\documentclass[12pt,preprint]{aastex}

\begin{document} 

\title{Followup Observations of SDSS and CRTS Candidate Cataclysmic 
Variables\altaffilmark{*}}

\author{Paula Szkody\altaffilmark{1,2},
Mark E. Everett\altaffilmark{3},
Steve B. Howell\altaffilmark{4,5,2},
Arlo U. Landolt\altaffilmark{6,2},
Howard E. Bond\altaffilmark{7,2},
David R. Silva\altaffilmark{3},
Stephanie Vasquez-Soltero\altaffilmark{1}} 

\altaffiltext{*}{Based on observations obtained with the Apache Point
  Observatory (APO) 3.5-meter telescope, which is owned and operated
  by the Astrophysical Research Consortium (ARC).}
\altaffiltext{1}{Department of Astronomy, University of Washington,
  Box 351580, Seattle, WA 98195; szkody@astro.washington.edu}
\altaffiltext{2}{Visiting Astronomer, Kitt Peak National Observatory, National
Optical Astronomy Observatory, which is operated by the Association of Universities 
for Reasearch in Astronomy (AURA) under cooperative agreement with the National Science 
Foundation}
\altaffiltext{3}{National Opical Astronomy Observatories, 950 N. Cherry Ave, Tucson, AZ 
85719; dsilva@noao.edu}
\altaffiltext{4}{NASA Ames Research Center, Moffett Field, CA 94035;steve.b.howell@nasa.gov}
\altaffiltext{5}{Visiting Astronomer, Mt. Palomar Observatory, Palomar Mountain, CA 92060}
\altaffiltext{6}{Department of Physics \& Astronomy, Louisiana State University, Baton Rouge, LA 70803; landolt@rouge.phys.lsu.edu}
\altaffiltext{7}{Department of Astronomy \& Astrophysics, Pennsylvania State University,
University Park, PA 16802; heb11@psu.edu}

\begin{abstract}

We present photometry of 11 and spectroscopy of 35 potential cataclysmic
variables from the Sloan Digital Sky Survey, the Catalina Real-Time Transient
Survey and vsnet-alerts. The photometry results include quasi-periodic oscillations during
the decline of V1363 Cyg, nightly accretion changes in the likely Polar (AM Herculis binary)
SDSS J1344+20, eclipses in SDSS J2141+05 with an orbital period of 76$\pm$2 min,
and possible eclipses in SDSS J2158+09 at an orbital period near 100 min.
Time-resolved spectra reveal short orbital periods near 80 min for
SDSS J0206+20, 85 min for SDSS J1502+33, and near 100 min for CSS J0015+26,
RXS J0150+37, SDSS J1132+62, SDSS J2154+15 and SDSS J2158+09. The prominent \ion{He}{2}
line and velocity amplitude of SDSS J2154+15 are consistent with a Polar nature
for this object, while the lack of this line and a low velocity amplitude
argue against this classification for RXS J0150+37. Single spectra of 10 objects
were obtained near outburst and the rest near quiescence, confirming the
dwarf novae nature of these objects.

\end{abstract}

\section{Introduction}

The observational identification of close binaries involving mass transfer from
a late main-sequence star to a white dwarf (cataclysmic variables or CVs) has
always been hampered by selection effects. The first wide field surveys like
the Palomar-Green (Green et al. 1982) concentrated on bright ($<$16 mag) blue-color
objects and so found many high accretion, novalike type systems. The Hamburg
Survey (Hagen et al. 1995) used low resolution spectra to identify emission line
objects (targeted for quasars) and was able to reach fainter ($<$17.5 mag)
 objects with strong
emission lines. The Sloan Digital Sky Survey (York et al. 2000) 
pushed to 21st-22nd magnitude in 5 filters
and to about 20th mag in medium resolution spectra and so was able to identify
the faintest, shortest period dwarf novae as well as extremely low accretion
systems with magnetic white dwarfs (summary in Szkody et al. 2011). While this
latter survey and its followup on individual objects provided a means to
counter previous biases due to brightness and provide a test of population
models (G\"ansicke et al. 2009), it suffered from incompleteness and a concentration
on CVs out of the galactic plane. 

Besides photometric colors and emission lines,
the other way to characterize CVs is by their variability, both long term via
outbursts and high/low states in novalikes, and short term via orbital variability
due to hot spots, eclipses, flickering (a summary of general properties of CVs
is contained in Warner 1995). This aspect is being pursued with the recent and ongoing 
Catalina Real-Time Transient
Survey, CRTS (Drake et al. 2009a) which uses 3 different telescopes: a 0.7m Schmidt
in the Catalina mountains (CSS) and a 1.5m Cassegrain on Mt. Lemmon (MLS) to cover
the northern skies, and a 0.5m Schmidt at Siding Springs (SSS) for the south. 
The CRTS also avoids the galactic plane but accomplishes long
term light curves down to about 20-21st mag. The candidate CVs 
are made public by the CRTS on their
website\footnote{crts.caltech.edu/}.
Two other all sky surveys that use small wide-field cameras but only reach
objects that brighten at outburst are the All Sky Automated Survey
(ASAS; Pojmanski (1997), and the Mobile Astronomical System of the Telescope
Robots, MASTER (Lipunov et al. 2010). Information about the CV systems discovered
in these surveys is usually dispersed via vsnet 
alerts\footnote{http://ooruri.kusastro.kyoto-u.ac.jp/pipermail/vsnet-alert/}.

While these surveys provide the database to identify CV candidates, the confirmation
and specific nature of the type of CV requires detailed followup observations using
spectroscopy and orbital lightcurves. These followup data determine whether an
accretion disk or the underlying stars contribute most of the light, whether
the system is a high excitation novalike showing \ion{He}{2}, if it contains a
magnetic white dwarf (Polar) showing cyclotron or Zeeman splitting features, and if the
the orbital period is long or short. This type of information for a large unbiased all sky
survey allows a comparison of observed numbers of CVs in each period range and each
category with expectations from population models.

Several groups are now accomplishing followup studies of the available candidate CVs.
 Woudt et al. (2012) and Coppejans et al. (2014)
 have published their photometry of 40 southern objects, Thorstensen
\& Skinner (2012; hereafter TS) have published spectra and photometric colors of 36
northern systems and Breedt et al. (2014; hereafter B14) accomplished spectroscopic
identification of 85 additional systems along with a discussion of the CRTS light 
curves. Since many of the candidates were discovered during outbursts, most are 
relatively faint
for followup at quiescence. TS and B14 compared the available results from CRTS to
the SDSS and Ritter-Kolb samples and concluded that the CRTS is biased toward large
outburst amplitudes, missing objects that are not dwarf novae. However, the large
number of new discoveries shows that there are still many CVs being missed.

Our group has continued to obtain followup photometry and spectroscopy
of both the SDSS sample and the CRTS and vsnet alerts 
to further the task of identifying the nature of the large number of available
CV candidates. Our results from 2010-2013 are presented here, which include
photometry of 11 CVs (8 from SDSS, one from CRTS and two from vsnet) and
spectroscopy of 35 (8 with SDSS spectra, 14 from CRTS that have only SDSS 
photometry, and 13 from CRTS, MASTER and ASAS that are not in the SDSS 
database). Time-resolved spectra of 8 objects were obtained, with tentative
fits to a sine-curve solution available for seven.  Our observations have 
six overlaps with
objects studied by TS: we have spectra of 3 objects
that TS have only photometric colors, while their spectral studies of the other
3 occurred during different years/outburst states. As done by the other groups,
for simplicity we 
identify all objects by their 2000 RA and Dec coordinates in Tables 1 and 2 
(which allows them
to be found in their respective databases) and abbreviate these coordinates in 
the following text discussion.

\section{Observations}

Most of the photometric observations were conducted on the KPNO 2.1 m telescope
during 2011 May and June. The STA2 CCD was used with either a V 
or a BG39 filter. Data were also acquired on three nights during 2013 July
and October with the UW 0.76m telescope at Manastash Ridge Observatory
 using a Spyder CCD with a BG40 filter. 
The data are summarized in Table 1. Reductions were
accomplished using IRAF\footnote{IRAF is distributed by the National
Optical Astronomy Observatory,
which is operated by the Association of Universities for Research in 
Astronomy, under cooperative agreement with the National Science 
Foundation.} routines under ${\it ccdproc}$ to flat-field and bias 
correct the
images. Magnitudes of the variables and comparison stars on the same
images were then measured using ${\it qphot}$ and used 
to construct differential light curves on each night. Due to relatively long
readout times and faintness of the targets, the time resolution of most of the light
curves is on the order of a minute.

Spectroscopy was primarily accomplished during 2010-2013 using both the KPNO 4 m
telescope and the 3.5 m telescope at Apache Point Observatory (APO).
At KPNO, the RC-Spectrograph was used with 
the 2048 CCD T2KA and a 1 arcsec slit. The second order of
grating KPC-22b produced a focussed spectrum over
3800-4900\AA\ with a resolution of 0.7\AA\ pixel$^{-1}$. FeAr
lamps were used to calibrate the wavelengths and flux standards were
observed. At APO, the Double-Imaging
Spectrograph was generally used in high resolution mode to provide
simultaneous blue and red spectral coverage with a resolution of
0.6\AA\ pixel$^{-1}$ for blue
wavelengths of 3900-5000\AA\ and red wavelengths of 6000-7200\AA. On one
night (2010 October 2), the low resolution gratings were used, resulting in a 
resolution of 1.2\AA\
from 3400-9500\AA. 
Flux standards and HeNeAr lamps were used for calibration. On one
other night (2010 September 18), the Double Beam Spectrograph on the 5m 
telescope at Mt. Palomar provided a spectrum with a resolving power of
7700 in the blue and 10000 in the red, with wavelength coverage of
3500-5000\AA\ in the blue and 6200-6800\AA\ in the red. Standards and
FeAr and HeNeAr lamps were used for calibration. In all cases, IRAF tasks
under ${\it ccdproc, apall}$ and ${onedspec}$ were used to correct
the images, extract the spectra to 1-d and calibrate them. 
For the time-resolved spectra, velocities of the Balmer emission lines
were measured with the
centroid ${\it e}$ routine in
${\it splot}$ and an IDL program was used to find the least-squares fit 
to a sine-curve. A summary of the spectroscopic observations is given in Table 2
and the radial-velocity fits are summarized in Table 3. A montage of the
available useful spectra is shown in Figure 1. As both the KPNO and APO
data cover the blue region of the spectrum, Figure 1 is plotted for the region
that is common to all spectra.

\section{Results for Systems with Light Curves}

The following sections provide the photometric and spectroscopic 
results obtained for the 11
 systems with observed light curves (Table 1).

\subsection{V1363 Cyg}

This CV has a peculiar nature which has defied classification.
It was classified as a dwarf nova when it was discovered (Miller 1971)
but the light curve was peculiar in showing long periods at quiescence
near 17.5 mag and also what appeared to be standstills about one magnitude below
maximum light that would classify it as a Z Cam star (Warner 1995). Bruch and Schimpke
(1992) obtained spectra which showed the typical Balmer emission lines of a CV.
Outbursts are relatively rare and have not been well-studied in the past.
In 2011 May, V1363 Cyg underwent its first outburst since 1952 as
reported by vsnet (Oshima 2011a). Quasi-periodic oscillations (QPOs)
 from 8.5-10.5 min
and amplitude of 0.05 mag were reported from May 8-June 1. On May 30,
a longer QPO at 28.8 min was observed (Oshima 2011b). Our light curves in May 
(Figure 2)
show the QPOs, with changes in amplitude noticeable in the longer lightcurves
of May 13 and 24. 

\subsection{RXS0150+37}

This counterpart to an x-ray object in Andromeda was found by MASTER and
reported by Denisenko 
(2013a) with archival images showing variability from 15-19th magnitude.
A further alert (Denisenko 2013b) reported unusual behavior at a mid-state
between 2013 Sept 16-25 that could
be indicative of an active Polar. Our 2.8 hrs of photometry on
2013 Oct 29 (Figure 3) show a double-humped modulation at a period of 109 min.
The velocities measured from our 1.5 hrs of APO time-resolved spectra on 2013
Oct 2 give a best fit to a sinusoid with period of 103 min and semi-amplitude
of 54 km s$^{-1}$ (Table 3 and Figure 3).
The lack of the high excitation line of \ion{He}{2}, combined with the low
semi-amplitude of the radial velocity curve argue for this object being
a normal low inclination dwarf nova rather than a Polar.

\subsection{SDSS1219+20}

This 19th magntitude CV was observed in order to search for possible 
white dwarf pulsations,
as its SDSS spectrum (Szkody et al. 2011) showed broad Balmer absorption lines
flanking the emission lines. This type of spectrum is typical in the 16 known
accreting white dwarf pulsators. However, our 4 light curves in 2011 May 
did not reveal
any periodic feature that could be ascribed to either pulsation or orbital
motion.

\subsection{SDSS1344+20}

The SDSS spectrum of this 17th magnitude 
object showed weak Balmer emission lines and a broad hump feature near 5200\AA\ 
that implied an origin as a cyclotron
harmonic from a magnetic white dwarf (Szkody et al. 2011). Time-resolved
spectra revealed a large ($\sim$400 km s$^{-1}$) semi-amplitude 
variation 
of the H$\alpha$ and H$\beta$ lines with an orbital period of 110 and 122
minutes respectively, which was consistent with a Polar nature. 
Our 3 nights
of photometry in 2011 June shown in Figure 4 corroborate this Polar identification, but
also show interesting changes during the week timespan. On June 5, a very large (2 magnitude)
hump is visible. This feature is typical of an accretion pole passing through the
line of sight such as occurs in the Polar VV Pup (Warner \& Nather, 1972; Liebert et al. 1978), 
but the light increases after the passage of the
pole. The next night, June 6, shows the hump with
less amplitude but longer duration, while 4 days later on June 10, the
system has transtioned to most of the time spent in the bright state. This
is somewhat reminiscent of the light curves of the 
Polar CSS071126+440405 (Thorne et al.
2010) which was
interpreted as sporadic mass transfer. If the increased mass transfer
moves the accretion zone so that it is no longer eclipsed by the white
dwarf, the light curve would not have the low segment visible on June 5.
The changes in the light curve make the determination of a photometric orbital
period difficult, but the features present on June 5 and 6 are most
consistent with a period near 100 min.

The spectra obtained in 2011 May and June (Table 2 and Figure 1) also reveal
significant changes in accretion. On May 15, the system was in a low state
with a featureless spectrum, while the spectra on June 9 and 10 show a
flux increase by a factor of 2 as well as increased Balmer line strengths
over those in the SDSS spectrum. If the broad hump around 4200\AA\ is a cyclotron
hump, its presence together with the one at 5300\AA\ in the SDSS spectrum
would imply a field strength of $\sim$65 MG for harmonics of 4 and 3
(Wickramasinghe \& Ferrario 2000).

\subsection{SDSS1604+41}

While the SDSS spectrum of this g=17.7 mag source shows strong \ion{He}{2} 
(Szkody et al. 2005), the light curve obtained over 1.4 hrs on 2013 July 26
shows no prominent feature, only flickering of a few tenths magnitude.  

\subsection{SDSS1605+20}

The SDSS spectrum of this 19.9 magnitude object (Szkody et al. 2009) shows
Balmer absorption surrounding its emission lines.  The 5 nights of photometry
in 2011 May-June (Figure 5) show a broad feature with an amplitude near 0.08 magnitudes and
a period of 75-78 minutes.

\subsection{SDSS1659+19}

This $g=16.8$ mag system also shows a prominent \ion{He}{2} line in its SDSS spectrum
(Szkody et al. 2006). Its light curve on 2 nights in 2013 July (the longest
timespan is shown in Figure 6) shows 10\% amplitude flickering as well as a
possible hump but further photometry is needed to resolve if this
is a periodic occurrence.

\subsection{SDSS2052-02}

This object was discovered in outburst at V=14 as ASASSN-13cg on 2013 Aug 27 and
matched with a blue SDSS g=18.3 object (Stanek 2013), while later vsnet posts suggested
a possible AM CVn object. Observations by Littlefield
reported by Kato (2013b) revealed a superhump period of 1.4 hrs
and shallow (0.1 mag) eclipses.
Our light curve 2 months later (2013 Oct 29) obtained near
quiescence (Figure 7) shows some variability near 37 min with an amplitude
of 0.4 mag although better S/N and a longer timespan will be needed to ascertain
if this is related to the orbital period.

\subsection{SDSS2141+05}

This $g=18.9$ mag CV appears in DR10 of the SDSSIII release. The SDSS spectrum
is shown in Figure 8. The deep absorption in the Balmer lines, along with
the broad absorption flanking the emission is indicative of a system at
high inclination and with a low mass transfer rate. Our followup photometry
during 2011 June (Figure 9) confirms the high inclination by showing 2 eclipses on each
night. While the eclipses are not well-resolved at our time resolution of
194 sec between exposures, the combination of the two nights yields an
orbital period of 76$\pm$2 minutes.

\subsection{CSS/SDSS2154+15}

An outburst of this CRTS CV on 2013 June 4 at 17.5 mag was reported by vsnet 
(Kato 2013a). A light curve showed a prominent single hump
feature with a period of 96.9 min. Our light curve 4 months later (Figure 10)
shows a similar feature. The large amplitude (1.5 mag) is indicative of
a Polar nature for this object. 

Our KPNO spectra obtained on 2013 Sept 5 (Figure 1) shows a prominent \ion{He}{2}
line. The velocities from the 82 min time-resolved spectra were fit to a
sine-curve with the period fixed at 96.9 min (Figure 10). The resulting
large K amplitudes near 300 km s$^{-1}$ (Table 3) are consistent with a polar.
The SDSS photometry obtained in 2009 October gives a fainter magnitude
($g=18.65$) and blue colors (u-g=0.10, g-r=0.02) which are not consistent with
the usual red colors of a polar. However, the system may have been in a low
state at the time of the SDSS observations, similar to the low state in EF Eri when
the white dwarf dominates the optical light (Wheatley \& Ramsay 1998).
Polarimetric measurements would confirm that this object harbors a magnetic white
dwarf.

\subsection{CSS/SDSS2158+09}

The CSS light curve of this object shows a bright measurement at 13.2 on
June 15, 2010. The rest of the time it resides near 17.6 mag, with a large
scatter indicative of orbital variability. This object exists in the SDSS
photometric database with blue colors (g=17.51, u-g=0.08, g-r=-0.10). Calibrated 
$V,R,I$ images by TS provide a comparable 
V magnitude of 17.48 and V-R=0.13. Our KPNO spectra obtained in
2010 Sept (Figure 1) show prominent deep central absorption in the Balmer emission 
lines indicative of a high inclination. Our KPNO photometry obtained 9 months later
(Figure 11) reveals a large amplitude (0.3 mag) hump feature followed by dips
(near UT times of 9 and 10.7 hr) which hint at eclipses but the 4 min integration 
times do not allow resolution of this 
feature. A periodogram of these data gives a period of 104 minutes. Using the
12 time-resolved spectra obtained over 2 hrs on 2010 Sept 16, we attempted a radial
velocity solution using centroids, gaussians and the line wings of H$\beta$.
Due to the complexity of the line shape and the less than ideal S/N, a good
solution could not be determined. The values obtained by fixing the period
at values between 100-120 min are listed in Table 3. The semi-amplitudes are
typical for dwarf novae but improved data will be needed to pin down the
period. Spectra in the red can determine if the secondary is visible.

\section{Time-Resolved Spectra of Systems without Light Curves}

In addition to the time-resolved spectra for the 3 objects with light curves described above, radial velocities
were also obtained for five other systems.

\subsection{CSS0015+26}

Drake et al. (2009b) identified this 18th mag ROSAT source as a CV when the CSS 
measured it at 13.3 mag in 2009 Sept. Our KPNO blue spectra over 73 min reveals
a typical dwarf nova (Figure 1) with a period near 100 min and a K semi-amplitude
near 70 km s$^{-1}$ (Figure 12, Table 3).

\subsection{CSS/SDSS0206+20}

The CSS identifies a one magnitude variation in this $g=15.58$ mag SDSS object. The blue
colors ($u-g=0.06,g-r=0.10$) are consistent with a CV. The KPNO blue
spectra obtained in 2013 Sept show the strong Balmer emission lines of a
dwarf nova and the 11 time-resolved spectra over 113 min on 6 Sept are fit
with a period near 80 min and a K semi-amplitude near 60 km s$^{-1}$ (Table 3,
Figure 13). While the photometric amplitude is very low, the other characteristics
are typical of a a dwarf nova. It is possible that the variation caught by
CSS was not an outburst but just a sporadic accretion event, as objects with
such short orbital periods often have outburst frequencies that are tens of years.

\subsection{SDSS1132+62}

This $g=18.5$ system showed strong Balmer emission lines in its SDSS spectrum
(Szkody et al. 2004). Our 82 min of time-resolved spectra on 2013 Feb 5
reveals a low K semi-amplitude and an orbital period near 100 min (Figure 14,
Table 3).

\subsection{SDSS1224+18}

The SDSS spectrum of this bright 16th mag system shows broad absorption with
central narrow emission and was suggested as a possible pre-CV (Szkody et al. 2011).Our 
126 min of time-resolved spectra on 2011 May 16 
showed no significant radial velocity variation
during that interval. The system likely has a long orbital period and/or a low
inclination. 

\subsection{SDSS/CSS1502+33}

This object was identified in the SDSS spectra at quiescence (Szkody et al. 2006) 
as a high inclination, eclipsing
system with a period of 84 min. A vsnet alert by Maehara (2011)
reported the CSS detection of this system at an outburst mag of 15 (2.5 mag
brighter than quiescence) on 2011 June 11. We obtained 4 spectra on June 13 and
8 on June 14 in order to search for a line component that could be from the
irradiated secondary star and yield the velocity amplitude of the secondary.
The spectra (Figure 1) show much stronger Balmer lines and a much larger flux in
 the \ion{He}{2} line than the quiescent spectra. The H$\beta$/\ion{He}{2} flux 
and equivalent width ratios on June 13 were 1.25 and 1.29 whereas they were
15 and 7 in the quiescent SDSS spectra. On June 13, the \ion{He}{2}
line does show a prominent red peak when the Balmer lines have a prominent
blue peak (Figure 1). However, the 4 spectra were not enough to resolve a
velocity curve and the next night, the peaks in the lines are similar in
phase. The radial velocity curves from June 14 (Table 3) have higher amplitudes
than the quiescent data (a factor of 2 for H$\beta$), implying an origin closer
to the white dwarf.

\section{Spectra at Outburst}

Ten of the objects listed in Table 2 and shown in Figure 1 were observed close to
their maximum reported brightness. Most of the ones near outburst peak 
show the typical
optically thick accretion disk that is the signature of outburst, with a rising
blue continuum and shallow Balmer absorption (ASAS0011+04, CSS0514+08, 
SDSS0754+38, SDSS0757+22, PNV1915+07, MASTER2355+42). Of these, only CSS0514+08
 shows the presence of 
\ion{He}{2}. TS have 3 spectra
from 4700-6700\AA\ for this object taken in 2011 January at quiescence which 
show H$\alpha$ and H$\beta$ in emission. Three other systems (CSS0353-03, 
SDSS0518-02
 and SDSS2100-02) have flat
or decreasing continua in the blue, which may be indicative of large reddening.
The presence of emission cores in the Balmer lines of SDSS0518-02 indicates the disk
is not at the peak outburst state.

The most interesting outburst spectrum exists for CSS0104-03. The CSS has
one recorded outburst of this system in 2009 but (Ohshima 2013) reported
an ASAS detection at 16.4 (about 3 mags above the SDSS and CSS quiescent
magnitudes) on 2013 Aug 27. The CSS lightcurve does not cover this period of 
time, but the large spread of magnitudes at quiescence indicates a possible 
eclipsing
system. Our KPNO spectra on 2013 Sep 03 show strong emission lines when
the system was still brighter than quiescence.

\section{Systems with Strong \ion{He}{2}}

Figure 1 shows 5 systems with a strong \ion{He}{2} emission line (SDSS0756+30, 
SDSS1232+22, SDSS1502+33,
SDSS1519+06, CSS2154+15). Previous sections have discussed SDSS1502+33 and
CSS2154+15. The remaining 3 systems were first presented in Szkody et al. 
(2011). Additonal spectra were taken to determine any changes that would
indicate the nature of these objects. All spectra are similar to the past
SDSS data, with weak broad Balmer and \ion{He}{2} emission lines.

\section{Comments on Remaining Spectra}

\subsection{CSS/SDSS0000+33}
CSS found an outburst of this object on 2010 Sept 10. Our KPNO spectrum on
Sept 12 shows the typical Balmer absorption lines of an accretion disk at outburst while
the Palomar spectrum on Sept 18 shows the lines in emission, indicating that it was
a short outburst and the system was already declining to quiescence. The SDSS
quiescent $g$ magnitude is 20.5. While the integration time is long, so there is likely
velocity smearing, the asymmetric nature of the lines may indicate a prominent hot
spot, but a larger telescope is needed for orbital resolution.

\subsection{CSS0051+20}
The spectrum of this system near quiescence shows an interesting sharp peak
in the broad Balmer emission lines. This could be indicative of irradiation of
the companion or a hot spot on the accretion disk.

\subsection{CSS/SDSS0058+28, CSS0650+41 and CSS1740+41}
While these objects are very faint at quiescence, broad lines of H$\beta$ and
H$\gamma$ are visible. The quiescent $g$ magnitude of SDSS0058+28 is 19.2.

\subsection{SDSS0344+09}

Szkody et al. (2011) identified this object as a possible CV but with
very narrow Balmer lines and a very blue continuum.  The KPNO spectrum is
very similar to the SDSS one, but extends blueward by 100\AA. It is
possible that this system is a hot star with an irradiated companion.

\subsection{CSS/SDSS0422+33}

The asymmetric line shape with a sharp narrow component is typical of
CVs with a prominent hot spot. The SDSS quiescent $g$ magnitude is 19.8.

\subsection{CSS0501+20}

 This object also displays the asymmetric profiles and peaks associated
with a hot spot component. TS have time-resolved red spectra obtained 4 months 
after our KPNO data from which they obtained an orbital period of 1.8 hr.

\subsection{CSS0647+49}

 TS obtained time-resolved red spectra in 2011 March that determined an orbital
period of 8.9 hrs. Our spectrum shows the blue wavelengths of this source.

\subsection{CSS/SDSS2156+19}
The spectrum shows prominent, very broad and doubled Balmer emission,
indicating a fairly high inclination system. TS obtained 4 images
on 2011 June at V=20.83 showing very blue colors (B-V=-0.96). The
SDSS photometry shows $g$=18.5.

\subsection{CSS/SDSS2243+08, CSS/SDSS2325-01, CSS/SDSS2338+28}
One year after their outbursts, these objects display typical CV spectra with 
strong, broad Balmer emission.
The SDSS photometry provides $g$ magnitudes of 19.6, 18.9 and 18.5 for these
3 objects. 

\section{Conclusions}

Our photometric and spectroscopic observations of CV candidates discovered from 
SDSS, CRTS and vsnet-alerts have resulted in the confirmation of objects as
dwarf novae, novalikes or Polars as well as the determination of unusual
properties of a few systems. Our photometry revealed QPOs following the outburst of
V1363 Cyg, eclipses in the short period system SDSS2141+05 and likely eclipses
in CSS/SDSS2158+09. Our spectra confirm the Polar nature of CSS/SDSS2154+15,
show interesting nightly changes in the accretion of the likely Polar
SDSS1344+20 and refute the Polar classification of RXS0150+37.
These results present systems where further detailed followup is necessary to
understand the physical parameters of the binaries. This work also
portends the impending problems in accomplishing the necessary followup
for future ever larger surveys such as LSST. It will be very difficult
to determine the precise nature of the CVs discovered without obtaining
both time-resolved spectra and photometry. A single spectrum at quiescence can confirm the
CV nature through the Balmer emission lines and the width of the lines and their single
or double-peaked nature gives information about the inclination while the strength of
\ion{He}{2} indicates a high excitation novalike or a magnetic white dwarf. However,
the time-resolved data are necessary for the determination of the orbital period through
the periodic presence of a hot spot or eclipses in the light curves, or its
determination through radial velocity curves which also provide the masses of the 
underlying stars. These physical parameters are necessary for formulating the correct
evolution and population model for close binaries. 

\acknowledgments
PS and SV-S acknowledge support from NSF grant AST-1008734. AL acknowledges 
NSF grant AST-0803158. H.E.B. thanks the STScI Director's Discretionary Research Fund for 
travel support. All authors thank the visionaries who created the National Observatory that
served the community so well for so many years.

\clearpage
\begin{deluxetable}{llllcccl}
\tabletypesize{\scriptsize}
\tablewidth{0pt}
\tablecaption{Summary of Photometric Observations}
\tablehead{
\colhead{UT Date} & \colhead{Object} & \colhead{Type,P(hr)\tablenotemark{a}} & 
\colhead{Obs} & \colhead{UT} & \colhead{Exp(sec)} 
& \colhead{Filter} & \colhead{State}}
\startdata
2011 May 12 & V1363 Cyg & DN & KPNO & 10:52-11:46 & 10 & BG39 & outburst \\
2011 May 13 & V1363 Cyg & DN & KPNO & 09:16-11:48 & 10 & BG39 & outburst \\
2011 May 24 & V1363 Cyg & DN & KPNO & 09:16-11:26 & 20 & BG39 & decline \\
2013 Oct 29 & CSS/SDSS215427+155938 & P,1.6 & MRO & 05:44-07:46 & 60 & BG40 & low \\
2011 Jun 28 & CSS/SDSS215815+094709 & DN,1.7 & KPNO & 07:59-11:21 & 240 & V & quies \\
2013 Oct 29 & RXS015017+375614 & DN,1.7 &MRO & 08:14-11:20 & 35 & BG40 & quies \\
2011 May 12 & SDSS121913+204938 & DN & KPNO & 04:33-06:06 & 30 & BG39 & quies \\
2011 May 13 & SDSS121913+204938 & DN & KPNO & 05:15-06:02 & 30 & BG39 & quies \\
2011 May 26 & SDSS121913+204938 & DN & KPNO & 04:49-05:36 & 30 & BG39 & quies \\
2011 May 27 & SDSS121913+204938 & DN & KPNO & 03:28-07:56 & 60 & BG39 & quies \\
2011 Jun 05 & SDSS134441+204408 & P,1.7 & KPNO & 04:00-06:25 & 30 & V & quies \\
2011 Jun 06 & SDSS134441+204408 & P,1.7 & KPNO & 03:35-05:59 & 30 & V & quies \\
2011 Jun 10 & SDSS134441+204408 & P,1.7 & KPNO & 03:46-07:12 & 30 & V & quies \\
2013 Jul 26 & SDSS160450+414328 & ... & MRO & 08:26-09:44 & 60 & BG40 & quies \\
2011 May 25 & SDSS160501+203056 & DN,1.3 & KPNO & 04:58-05:3  & 50 & BG39 & quies \\
2011 May 25 & SDSS160501+203056 & DN,1.3 & KPNO & 08:01-10:02 & 50 & BG39 & quies \\
2011 May 26 & SDSS160501+203056 & DN,1.3 & KPNO & 07:51-09:58 & 50 & BG39 & quies \\
2011 May 27 & SDSS160501+203056 & DN,1.3 & KPNO & 08:13-09:54 & 50 & BG39 & quies \\
2011 Jun 08 & SDSS160501+203056 & DN,1.3 & KPNO & 07:43-08:53 & 50 & BG39 & quies \\
2011 Jun 10 & SDSS160501+203056 & DN,1.3 & KPNO & 07:27-09:28 & 50 & BG39 & quies \\
2013 Jul 26 & SDSS165951+192745 & ... & MRO & 06:07-08:14 & 60 & BG40 & quies \\
2013 Jul 28 & SDSS165951+192745 & ... & MRO & 06:09-07:49 & 60 & BG40 & quies \\
2013 Oct 29 & SDSS205252-023952 & E,1.4 & MRO & 03:44-05:22 & 60 & BG40 & quies \\
2011 Jun 29 & SDSS214140+050729 & E,1.3 & KPNO & 09:00-11:20 & 60 & V & quies \\
2011 Jun 30 & SDSS214140+050729 & E,1.3 & KPNO & 07:19-09:15 & 60 & V & quies \\  
\enddata
\tablenotetext{a}{Provisional type of Dwarf Nova (DN), Polar (P), Eclipsing (E), and orbital period if determined}
\end{deluxetable}

\clearpage
\begin{deluxetable}{lllcccll}
\tabletypesize{\scriptsize}
\tablewidth{0pt}
\tablecaption{Summary of Spectroscopic Observations}
\tablehead{
\colhead{UT Date} & \colhead{Coords} & \colhead{Type,P(hr)\tablenotemark{a}} & 
\colhead{Source} & \colhead{Obs} 
& \colhead{UT start} & \colhead{Exp(sec)} & \colhead{State}}
\startdata
2010 Sep 12 & 000025+332543 & DN & CRTS,SDSS & KPNO & 08:27 & 900 & outburst \\
2010 Sep 18\tablenotemark{b} & 000025+332543 & DN & CRTS,SDSS & Pal & 09:26 & 1200 & decline \\
2013 Sep 03\tablenotemark{b} & 001133+045122 & DN & ASASSN-13ck & KPNO & 09:38 & 900 & outburst \\
2013 Oct 03 & 001133+045122 & DN & ASASSN-13ck & APO & 04:37 & 600 & outburst \\
2010 Sep 14 & 001538+263657 & DN,1.7 & CRTS & KPNO & 09:04 & 900 & quies \\
2011 Aug 26\tablenotemark{b} & 001538+263657 & DN,1.7 & CRTS & KPNO & 11:10 & 900x3 & quies \\
2011 Aug 29 & 001538+263657 & DN,1.7 & CRTS & KPNO & 10:29 & 600x8 & quies \\
2010 Sep 15 & 005153+204017 & DN & CRTS & KPNO & 08:27 & 1200 & quies \\
2013 Jan 16 & 005825+283004 & DN & CRTS,SDSS & APO & 02:01 & 900 & quies \\
2013 Sep 03 & 010411-031341 & DN & CRTS,SDSS & KPNO & 10:45 & 1200x2 & outburst \\
2013 Oct 03 & 015017+375614 & DN,1.7 & MASTER & APO & 04:55 & 600x8 & high \\
2013 Sep 01 & 020633+205707 & DN,1.3 & CRTS,SDSS & KPNO & 11:01 & 900 & out \\
2013 Sep 05 & 020633+205707 & DN,1.3 & CRTS,SDSS & KPNO & 11:06 & 600x5 & out \\
2013 Sep 06\tablenotemark{b} & 020633+205707 & DN,1.3 & CRTS,SDSS & KPNO & 09:39 & 600x11 & out \\
2010 Sep 15 & 034420+093006 & ... & SDSS & KPNO & 09:01 & 1200 & quies \\
2013 Jan 16 & 035318-034847 & DN & CRTS & APO & 03:13 & 900 & outburst \\
2013 Jan 16 & 042218+334215 & DN & CRTS,SDSS & APO & 04:41 & 900x3 & decline \\
2010 Sep 12 & 050124+203818 & DN,1.8 & CRTS & KPNO & 11:40 & 900 & quies \\
2013 Jan 16 & 051458+083503 & DN & CRTS & APO & 04:12 & 600 & outburst \\
2013 Sep 01 & 051815-024503 & DN & CRTS & KPNO & 11:51 & 900 & outburst \\
2010 Sep 12 & 064729+495027 & DN,8.9 & CRTS & KPNO & 11:00 & 900 & quies \\
2013 Sep 03 & 065037+413053 & DN & CRTS & KPNO & 11:39 & 750 & quies \\
2013 Feb 05 & 075418+381225 & DN & CRTS,SDSS & APO & 02:27 & 600x2 & outburst \\
2011 May 15 & 075648+305805 & ... & SDSS,CRTS & KPNO & 03:38 & 600x2 & quies \\
2013 Apr 01 & 075713+222253 & DN & CRTS,SDSS & APO & 03:40 & 600  & outburst \\
2013 Feb 05 & 113215+624900 & DN,1.7 & SDSS & APO & 04:39 & 600x8 & quies \\
2011 May 16 & 122405+184102 & ... & SDSS & KPNO & 03:50 & 600x12 & quies \\
2011 Jun 12 & 123255+222209 & ... & SDSS & KPNO & 03:56 & 900x3 & quies \\
2011 May 15 & 134441+204408 & P,1.7 & SDSS & KPNO & 04:40 & 900x3 & low \\
2011 Jun 09\tablenotemark{b} & 134441+204408 & P,1.7 & SDSS & KPNO & 04:14 & 900 & high \\
2011 Jun 10 & 134441+204408 & P,1.7 & SDSS & KPNO & 04:06 & 900x2 & high \\
2011 Jun 13\tablenotemark{b} & 150240+333423 & DN,1.4 & SDSS,CRTS & KPNO & 04:09 & 600x4 & decline \\
2011 Jun 14 & 150240+333423 & DN,1.4 & SDSS,CRTS & KPNO & 03:48 & 480x8 & decline \\
2011 May 16 & 151915+064529 & ... & SDSS & KPNO & 03:28 & 600 & quies \\
2011 Jun 11\tablenotemark{b} & 151915+064529 & ... & SDSS & KPNO & 03:49 & 600x5 & quies \\
2013 May 05 & 174033+414756 & DN & CRTS & APO & 05:49 & 600x5 & outburst \\
2013 Jun 12 & 174033+414756 & DN & CRTS & APO & 05:52 & 900 & decline \\
2013 Oct 03\tablenotemark{b} & 174033+414756 & DN & CRTS & APO & 01:59 & 900x2 & quies \\
2013 Jun 12 & 191501+071847 & DN & PNV & APO & 06:38 & 300 & outburst \\
2013 Oct 03 & 210016-024258 & DN & CRTS,SDSS & APO & 04:13 & 900 & outburst \\
2013 Sep 01\tablenotemark{b} & 215427+155713 & P,1.6 & CRTS,SDSS & KPNO & 09:49 & 1200 & out \\
2013 Sep 05 & 215427+155713 & P,1.6 & CRTS,SDSS & KPNO & 09:18 & 600x8 & out \\
2010 Oct 02 & 215636+193242 & DN & CRTS,SDSS & APO & 03:26 & 900 & quies \\
2010 Sep 14\tablenotemark{b} & 215815+094709 & DN,1.7 & CRTS,SDSS & KPNO & 08:26 & 900x2 & quies \\
2010 Sep 16 & 215815+094709 & DN,1.7 & CRTS,SDSS & KPNO & 07:07 & 600x12 & quies \\
2010 Oct 02 & 224348+080927 & DN & CRTS,SDSS & APO & 03:48 & 900 & quies \\
2010 Oct 02 & 232551-014024 & DN & CRTS,SDSS & APO & 05:17 & 600 & quies \\
2010 Sep 12 & 233849+281955 & DN & CRTS,SDSS & KPNO & 10:18 & 900 & decline \\
2013 Sep 03 & 235503+420010 & DN & MASTER & KPNO & 09:57 & 1200x2 & outburst \\
\enddata
\tablenotetext{a}{Provisional type of Dwarf Nova (DN), Polar (P), Eclipsing (E),and orbital period if determined}
\tablenotetext{b}{For multiple observations, denotes spectrum shown in Figure 1}
\end{deluxetable}
\clearpage

\begin{deluxetable}{llcccc}
\tablewidth{0pt}
\tablecaption{Radial Velocity Fits}
\tablehead{
\colhead{Object} & \colhead{Line} & \colhead{P(min)} & 
\colhead{$\gamma$(km s$^{-1}$)} & \colhead{K (km s$^{-1}$)} &
\colhead{$\sigma$ (km s$^{-1}$)} } 
\startdata
001538+26 & H$\beta$ & 103 & -66$\pm$3 & 71$\pm$10 & 19 \\
001538+26 & H$\gamma$ & 99 & -53$\pm$2 & 78$\pm$9 & 16 \\
015017+37 & H$\alpha$ & 103 & 96.9$\pm$0.4 & 54$\pm$3 & 5.5\\
015017+37 & H$\beta$ & 103\tablenotemark{a} & 88$\pm$1 & 38$\pm$11 & 21 \\
020633+20 & H$\beta$ & 81 & -58$\pm$2 & 58$\pm$8 & 18 \\
113215+62 & H$\alpha$ & 100\tablenotemark{a} & -0.7$\pm$0.5 & 34$\pm$6 & 11 \\
113215+62 & H$\beta$ & 100\tablenotemark{a} & -7.2$\pm$0.5 & 20$\pm$7 & 12 \\
113215+62 & H$\gamma$ & 102 & -4.8$\pm$0.8 & 29$\pm$9 & 15 \\
150240+33 & H$\beta$ & 84.83\tablenotemark{a} & 6$\pm$2 & 104$\pm$26 & 36 \\
150240+33 & \ion{He}{2} & 84.83\tablenotemark{a} & 13$\pm$1 & 142$\pm$14 & 19 \\
215427+15 & \ion{He}{2} & 97 & -44.4$\pm$0.1 & 311$\pm$10 & 17 \\
215427+15 & H$\beta$ & 97\tablenotemark{a} & -2$\pm$1 & 293$\pm$51 & 89 \\
215815+09 & H$\beta$ & 100-120\tablenotemark{a} & -4--14 & 71-76 & 27-21 \\ 
\enddata
\tablenotetext{a}{Period fixed at this value}
\end{deluxetable}

\clearpage

\begin{figure}
\figurenum {1a}
\plotone{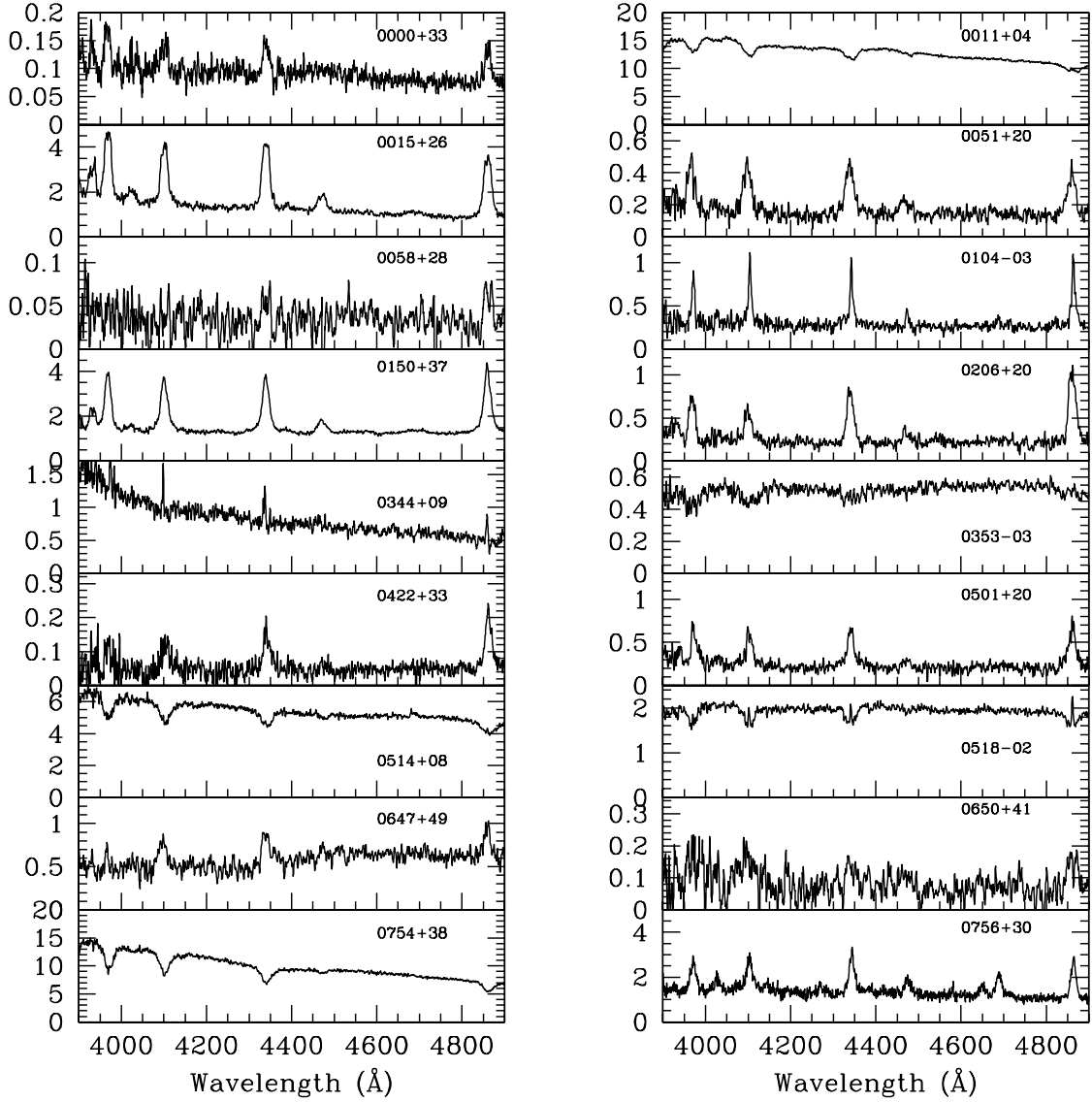}
\caption{Blue region spectra of sources listed in Table 2. Vertical axes are
F$_{\lambda}$ in units of 10$^{-15}$ ergs cm${-2}$ s$^{-1}$ \AA$^{-1}$.
Objects are labelled with first digits of RA and Dec as given in Table 2.}
\end{figure}

\clearpage

\begin{figure}
\figurenum {1b}
\plotone{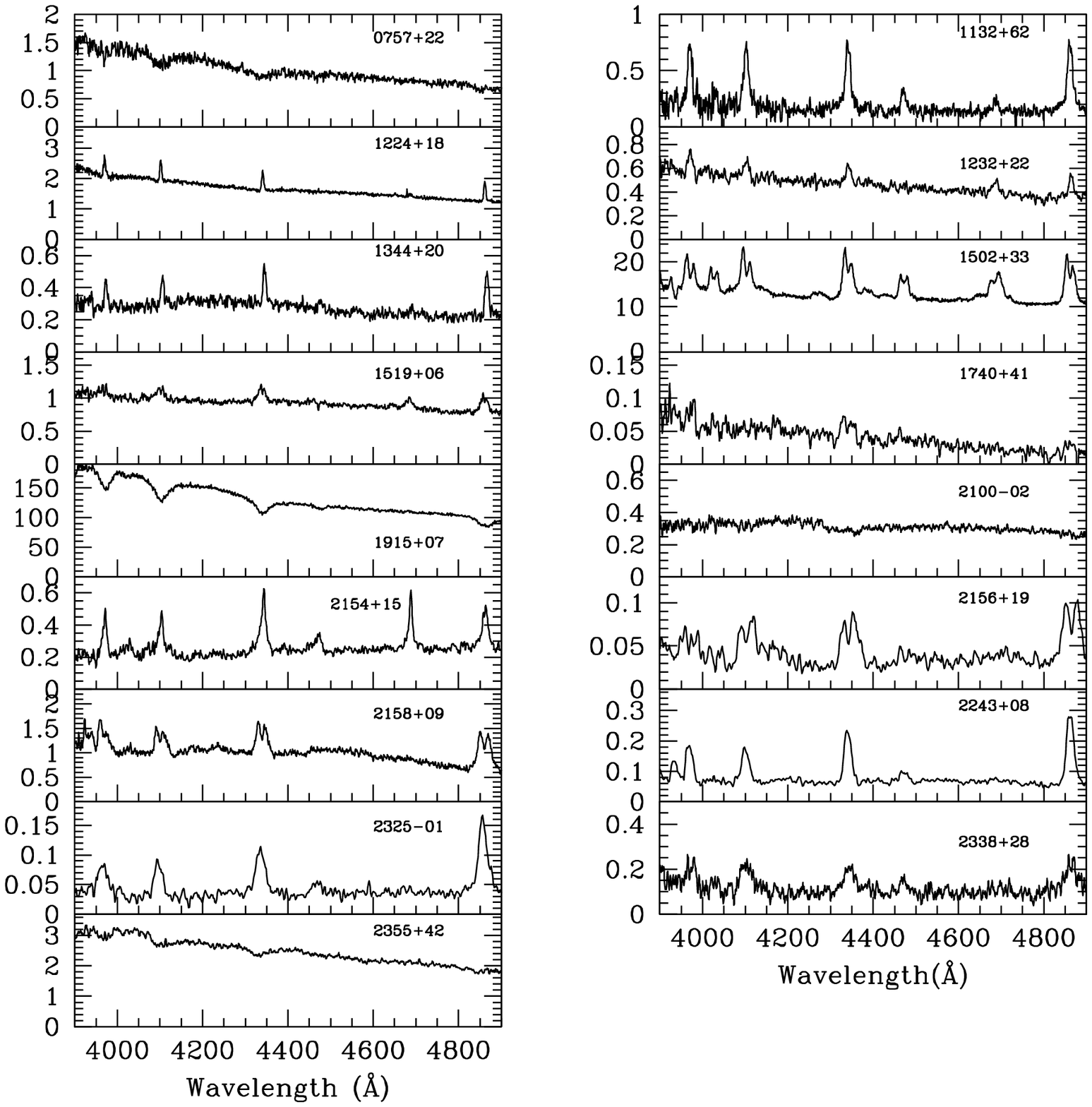}
\caption{Figure 1 continued.}
\end{figure}

\begin{figure}
\figurenum {2}
\plotone{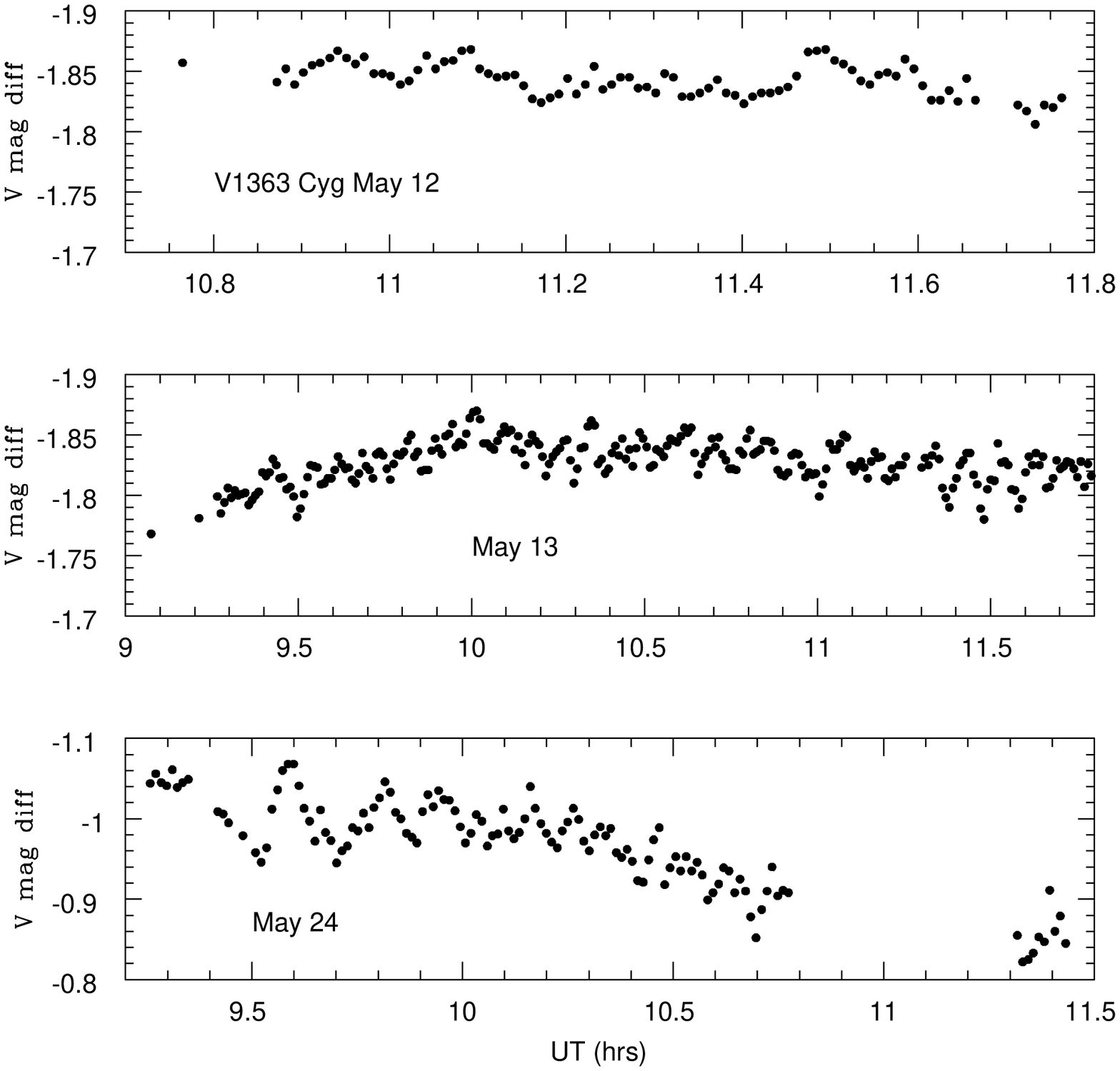}
\caption{KPNO 2011 May data showing the changing oscillations of the dwarf nova 
V1363 Cyg as the
system declined from outburst.}
\end{figure}

\clearpage
\begin{figure} 
\figurenum {3}
\epsscale{0.8}
\plotone{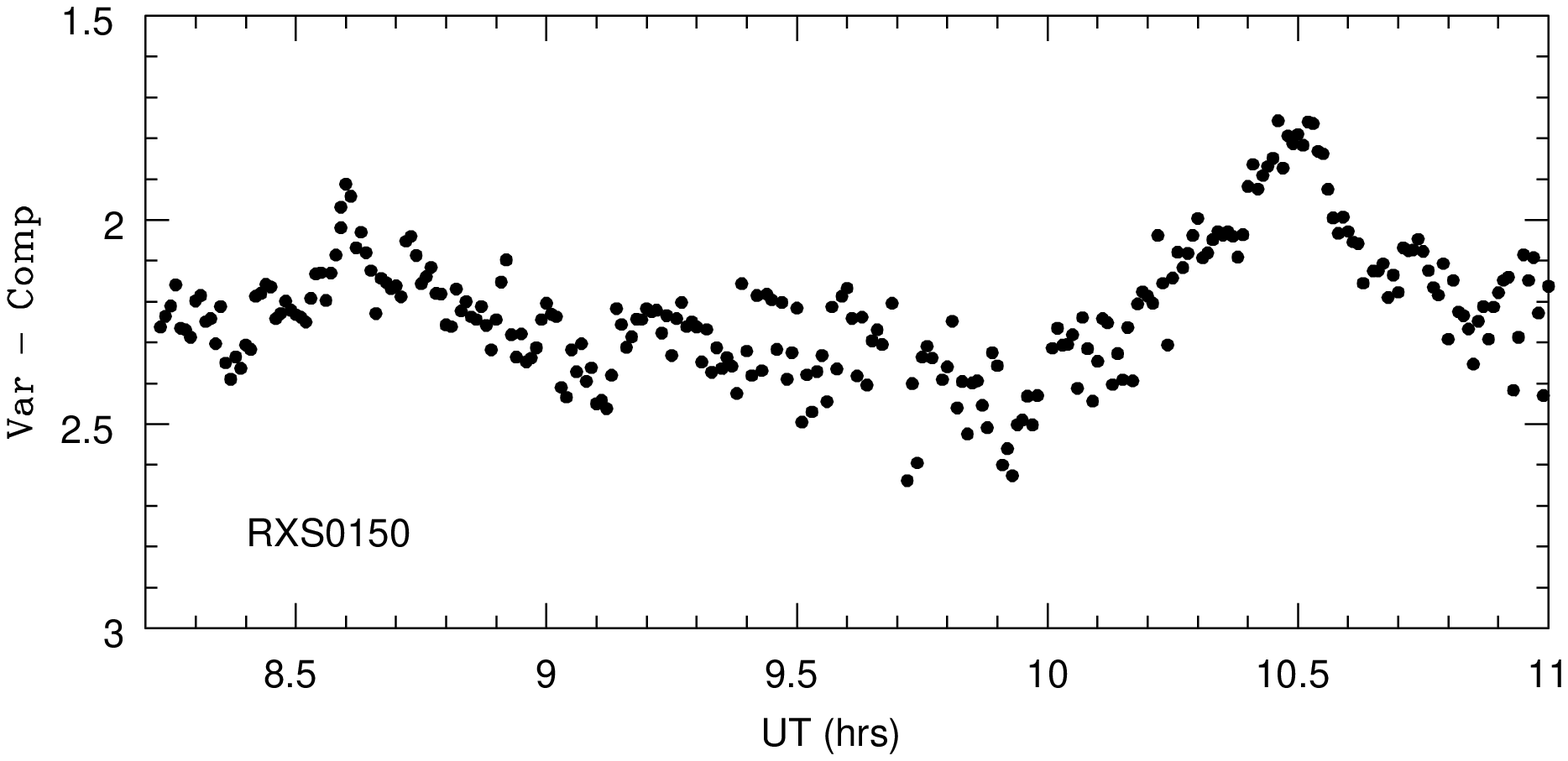}
\vspace{-2in}
\plotone{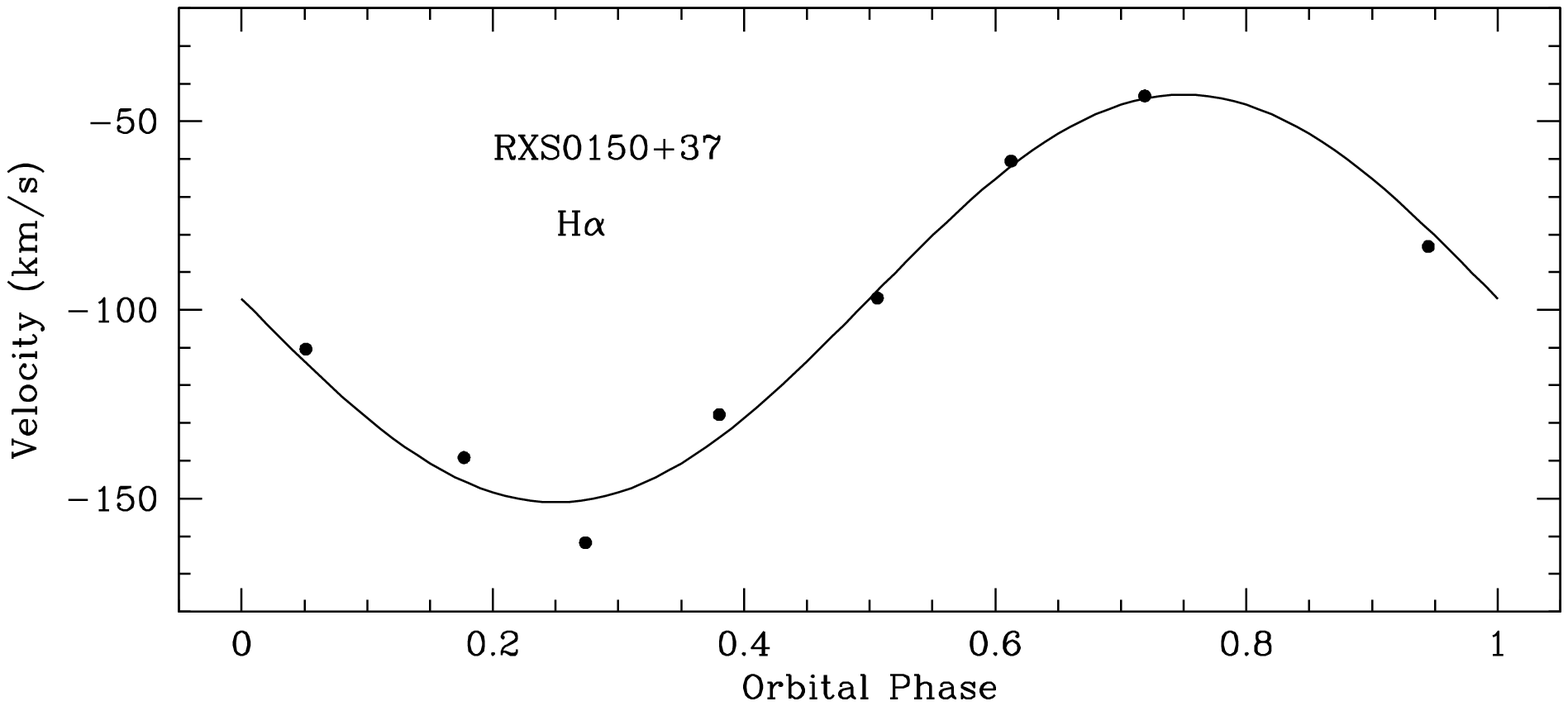}
\caption{Light curve of the dwarf nova RXS0150+37 on 2013 Oct. 29 (top) and radial velocity
curve (bottom) from 2013 Oct. 3. Velocity curve is phased with a period of
103 min and best fit sine curve is shown with parameters from Table 3.}
\end{figure}

\clearpage
\begin{figure}
\figurenum {4}
\plotone{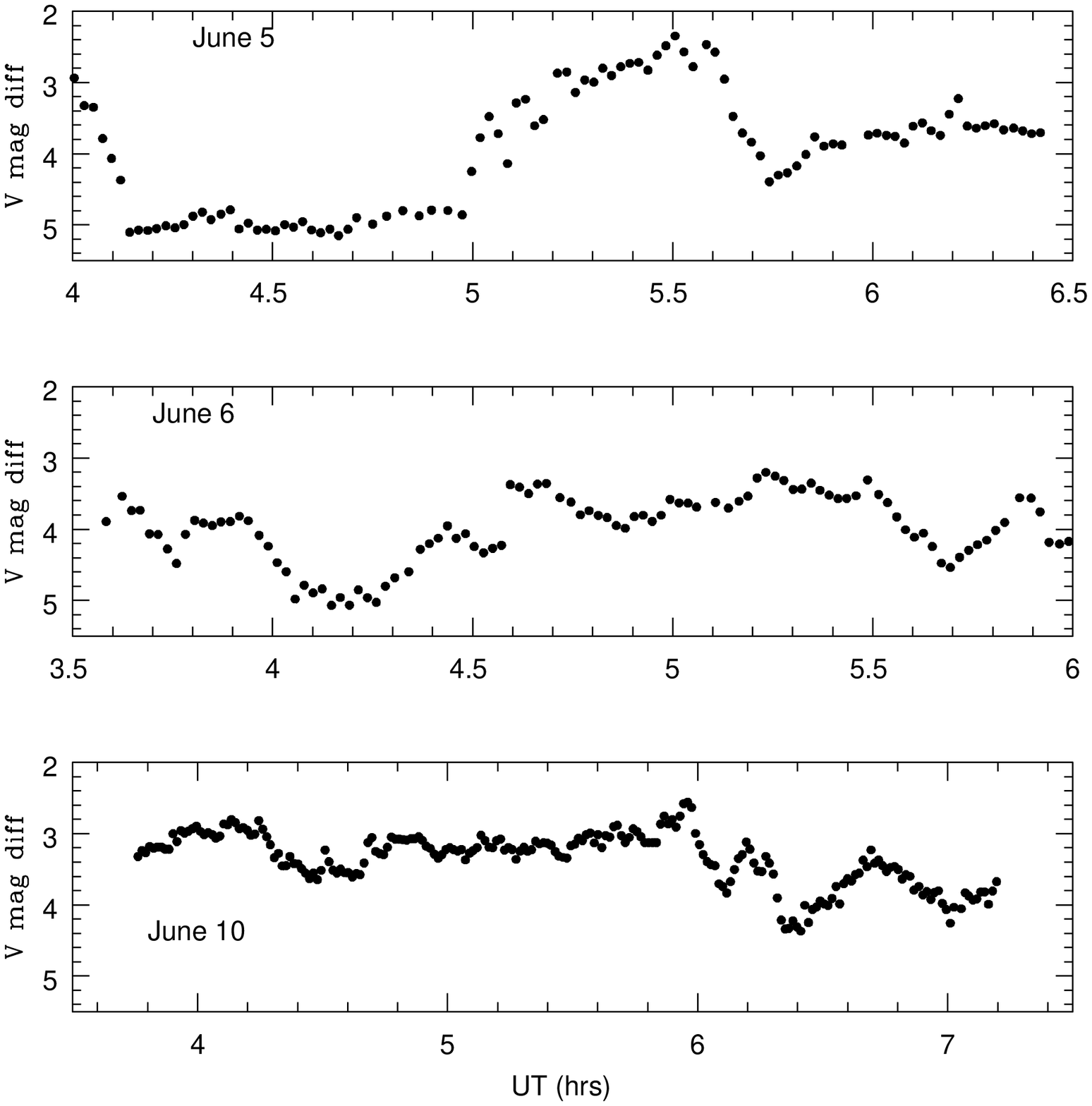}
\caption{KPNO data from 2011 June showing the large changes in the light
curve of the likely Polar SDSS1344+20 during the course of a week.}
\end{figure}

\clearpage
\begin{figure}
\figurenum {5}
\plotone{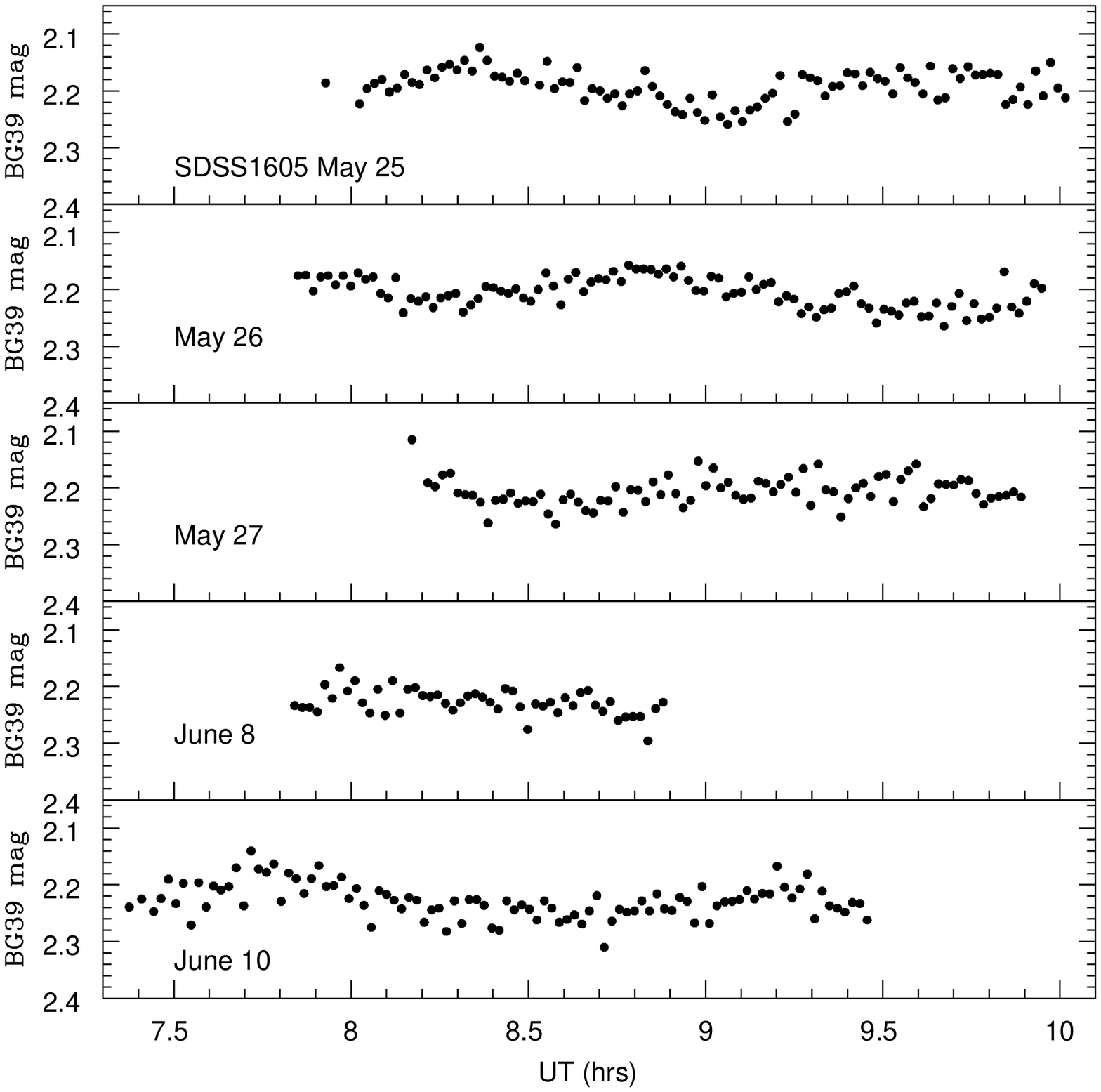}
\caption{KPNO data from 2011 May-June showing the 76 min orbital modulation
 in the light curves of the dwarf nova SDSS1605+20.}
\end{figure}

\clearpage
\begin{figure}
\figurenum {6}
\plotone{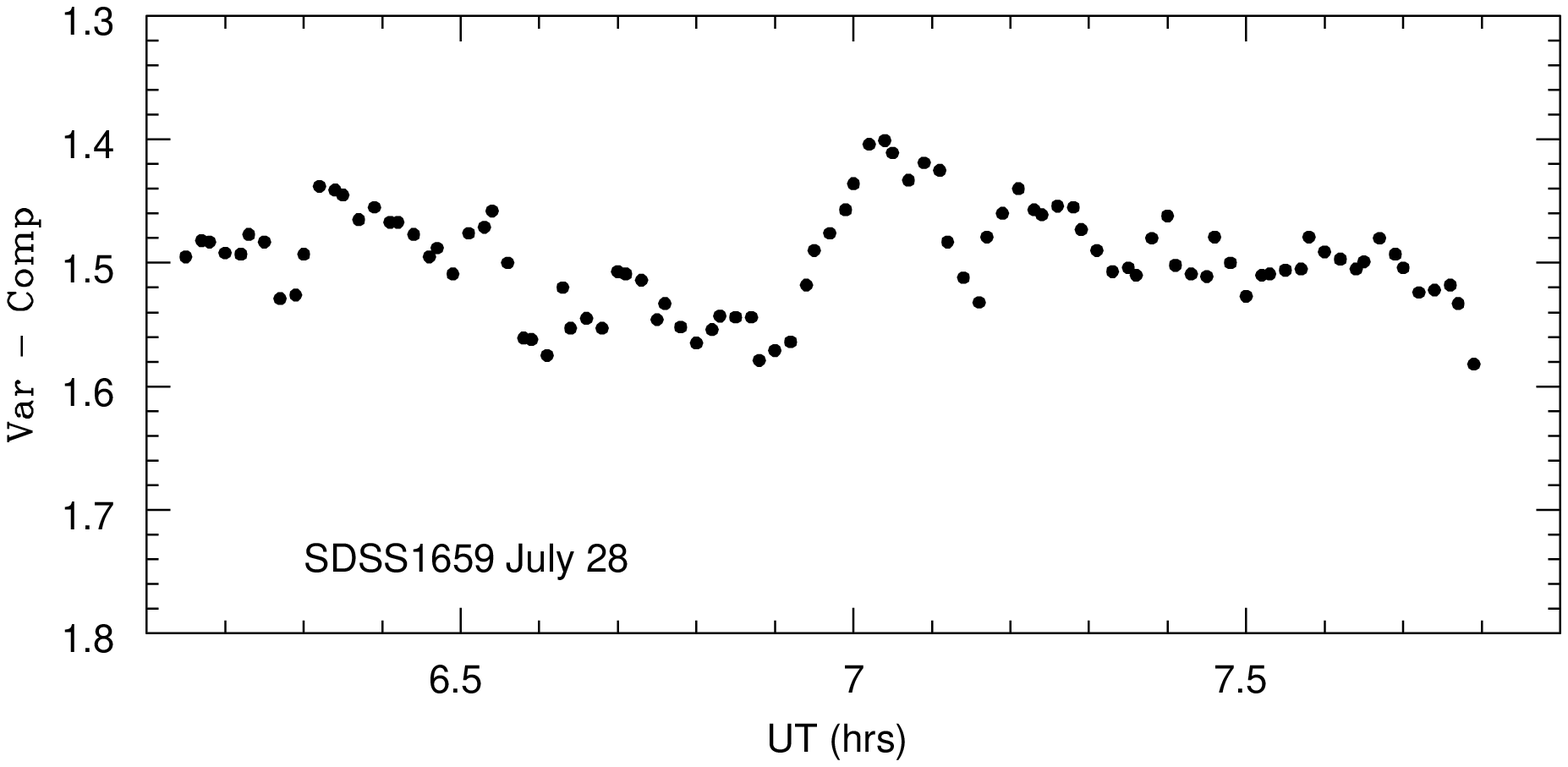}
\caption{MRO light curve of the suggested AM CVn system 
SDSS1659+19 obtained 2013 July 28.}
\end{figure}

\clearpage
\begin{figure}
\figurenum {7}
\plotone{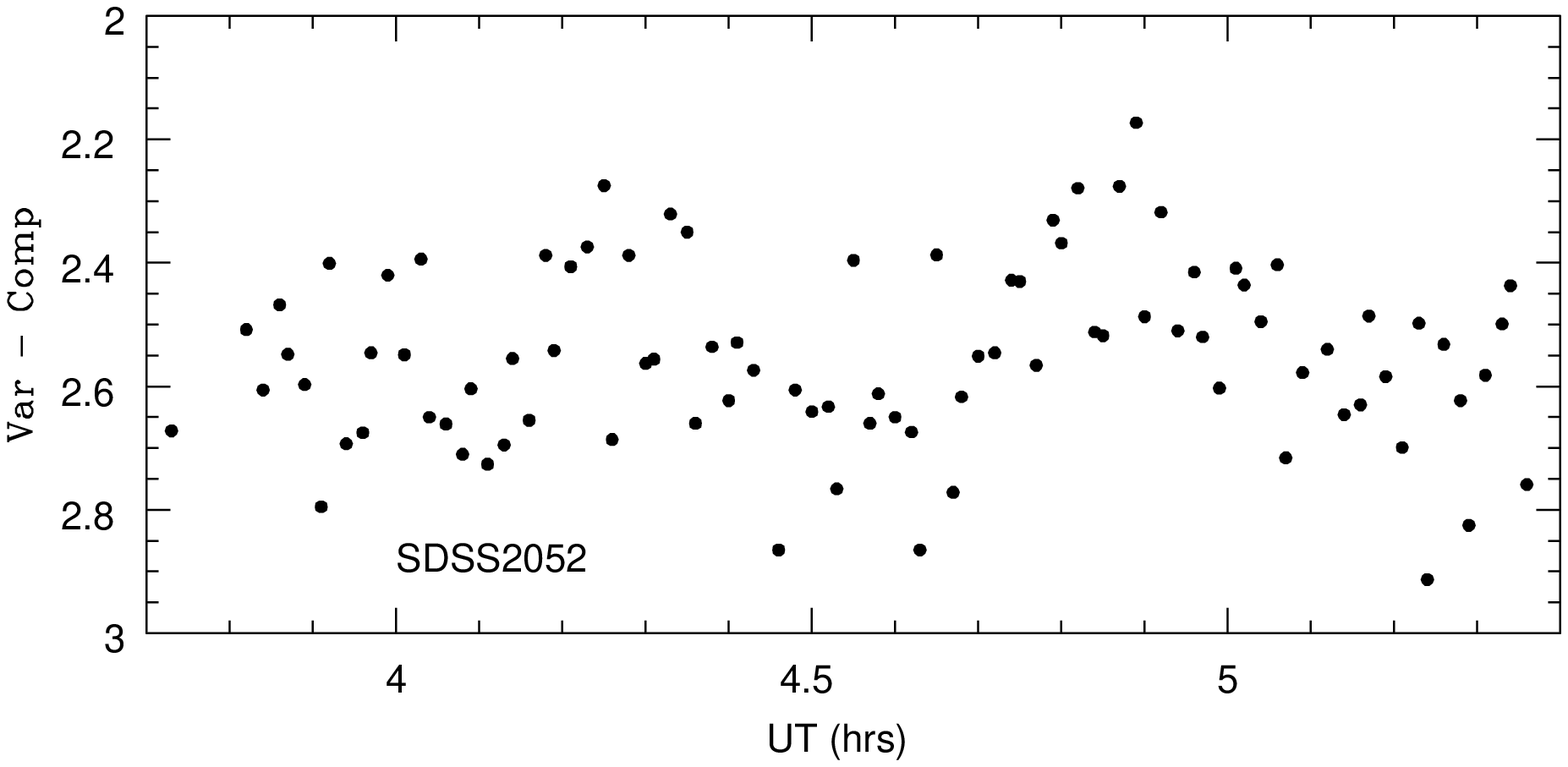}
\caption{MRO light curve of SDSS2052-02 obtained 2013 October 29.}
\end{figure}

\clearpage
\begin{figure}
\figurenum {8}
\plotone{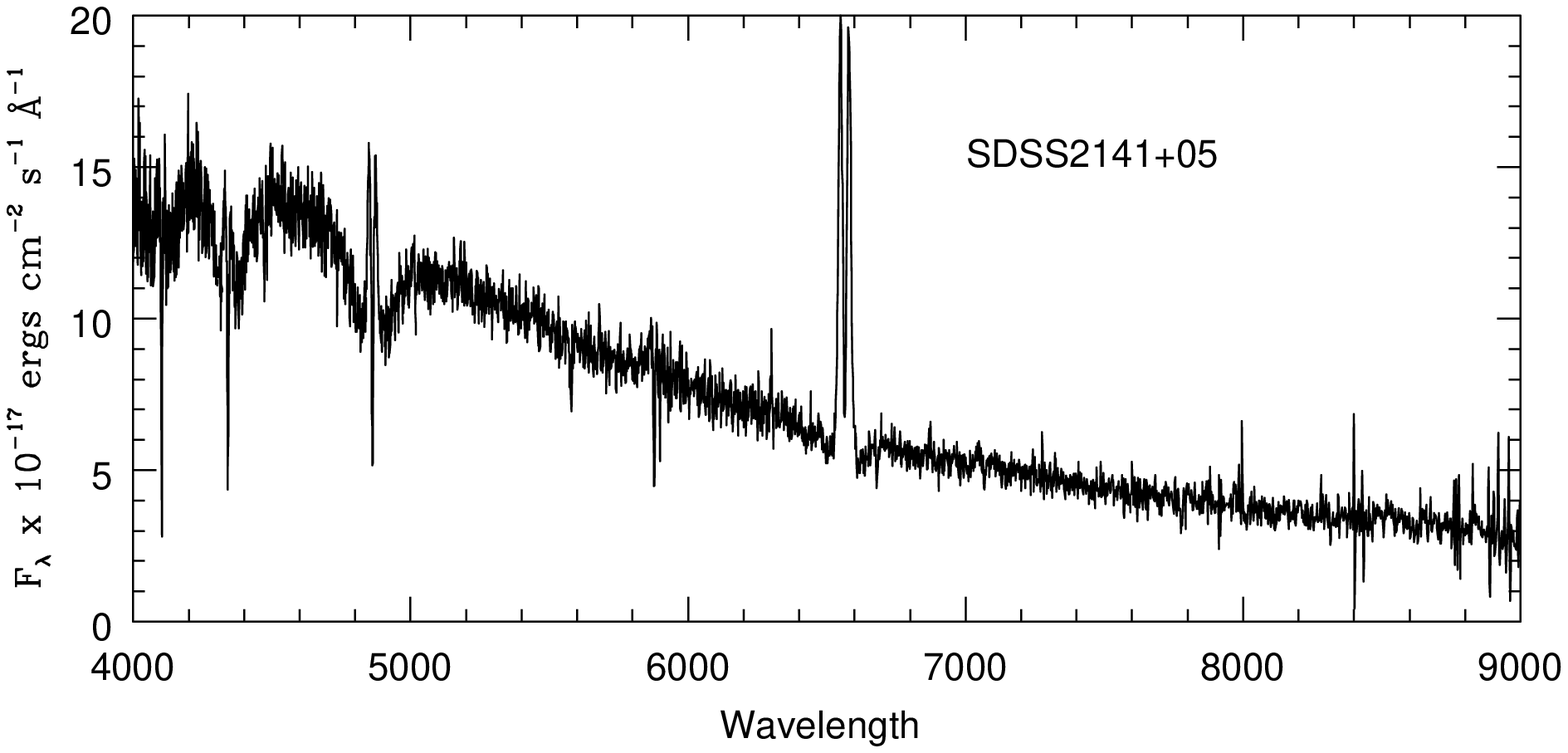}
\caption{SDSS spectrum from DR10 showing the prominent doubled emission lines
with central and flanking absorption indicative of a high inclination CV.}
\end{figure}

\clearpage
\begin{figure}
\figurenum {9}
\plotone{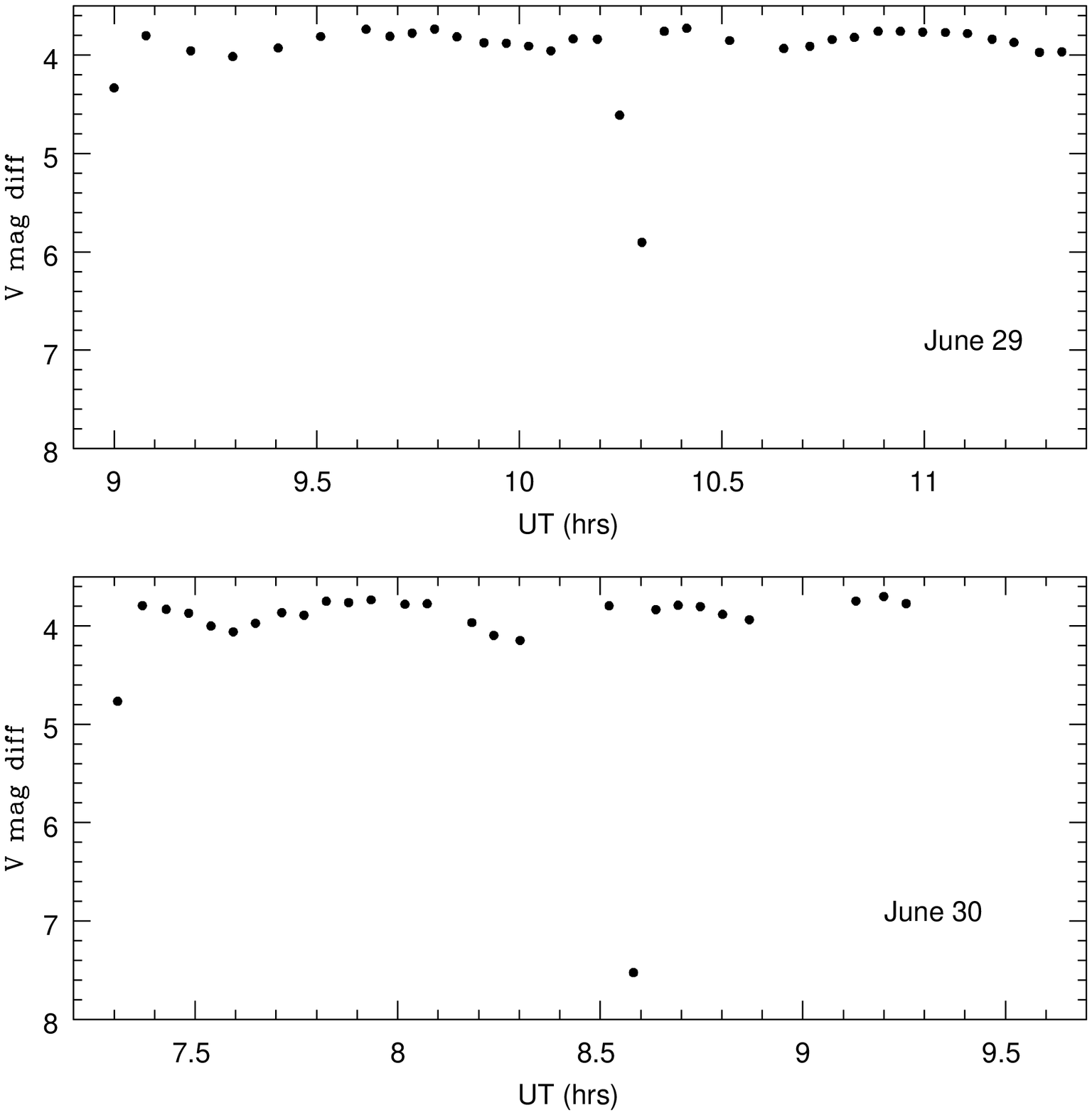}
\caption{KPNO data from 2011 June showing 2 eclipses and 2 egresses from eclipse in
 SDSS2141+05.}
\end{figure}

\clearpage
\begin{figure}
\figurenum {10}
\epsscale{0.8}
\plotone{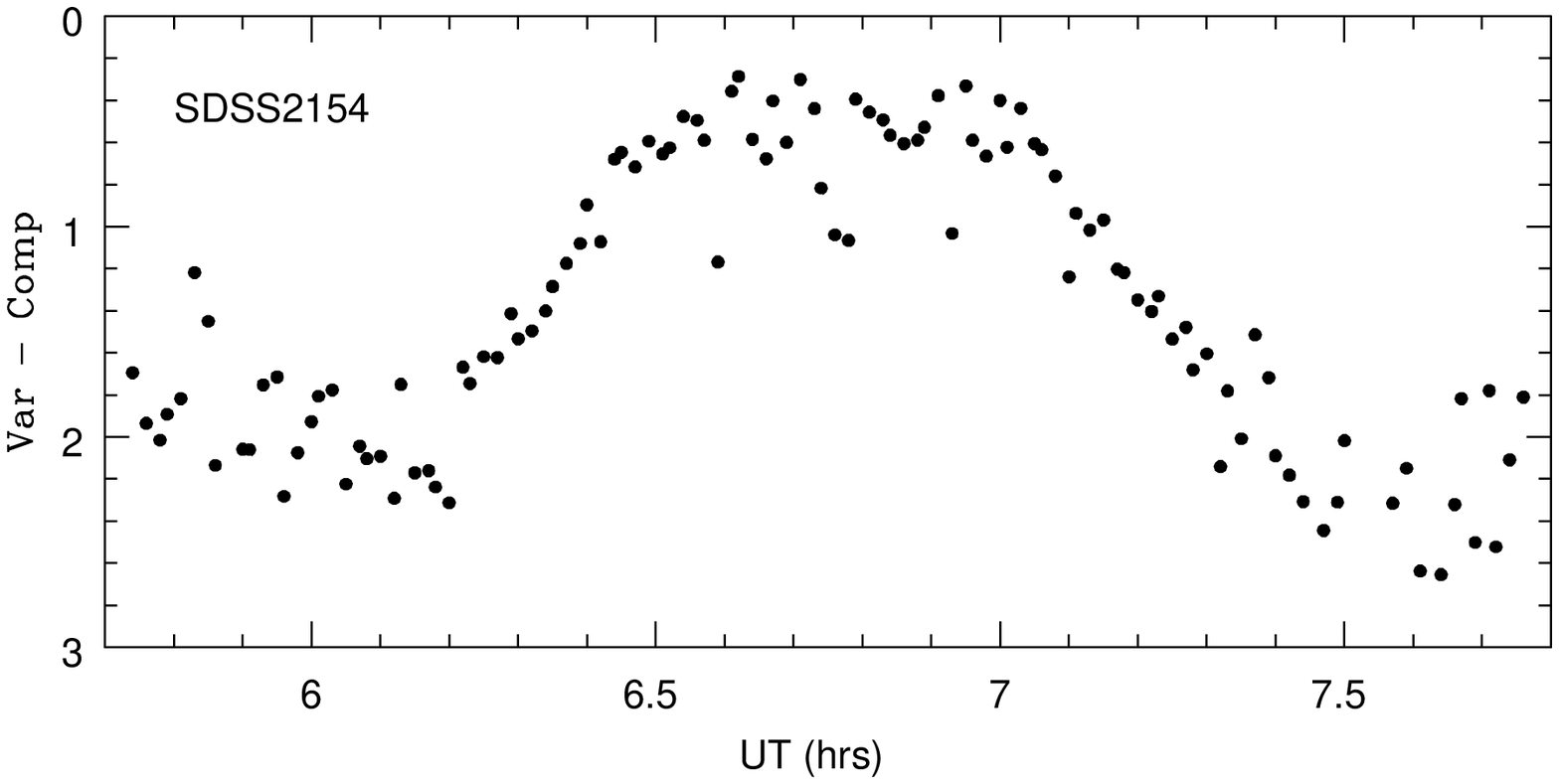}
\vspace{-2in}
\plotone{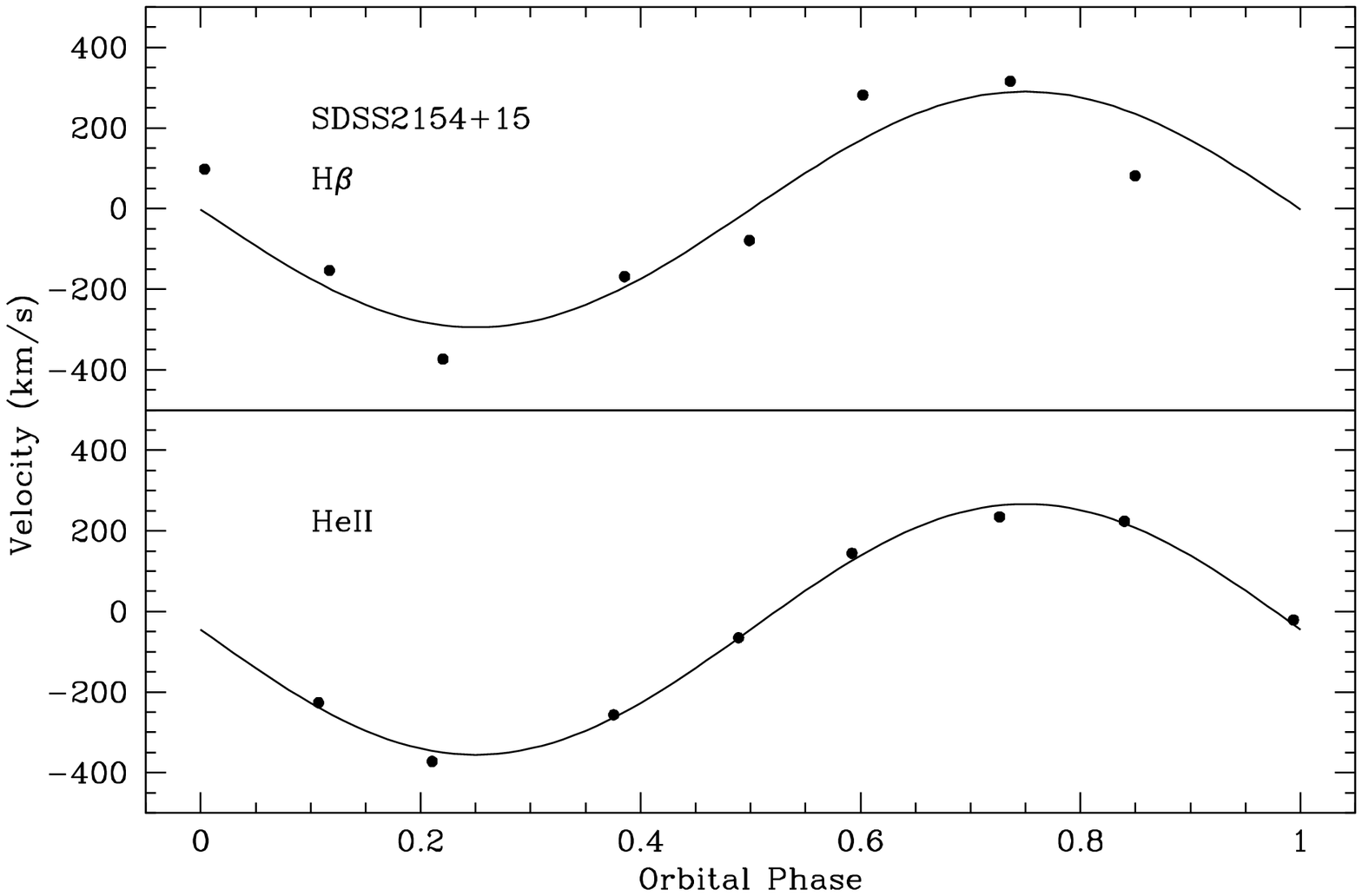}
\caption{MRO light curve of the likely Polar 
CSS/SDSS2154+15 obtained 2013 October 29 (top)
and the velocity curve from 2013 September 5 (bottom), phased with a period
of 97 min and showing the best-fit sine curve with parameters from Table 3.}
\end{figure}

\clearpage
\begin{figure}
\figurenum {11}
\plotone{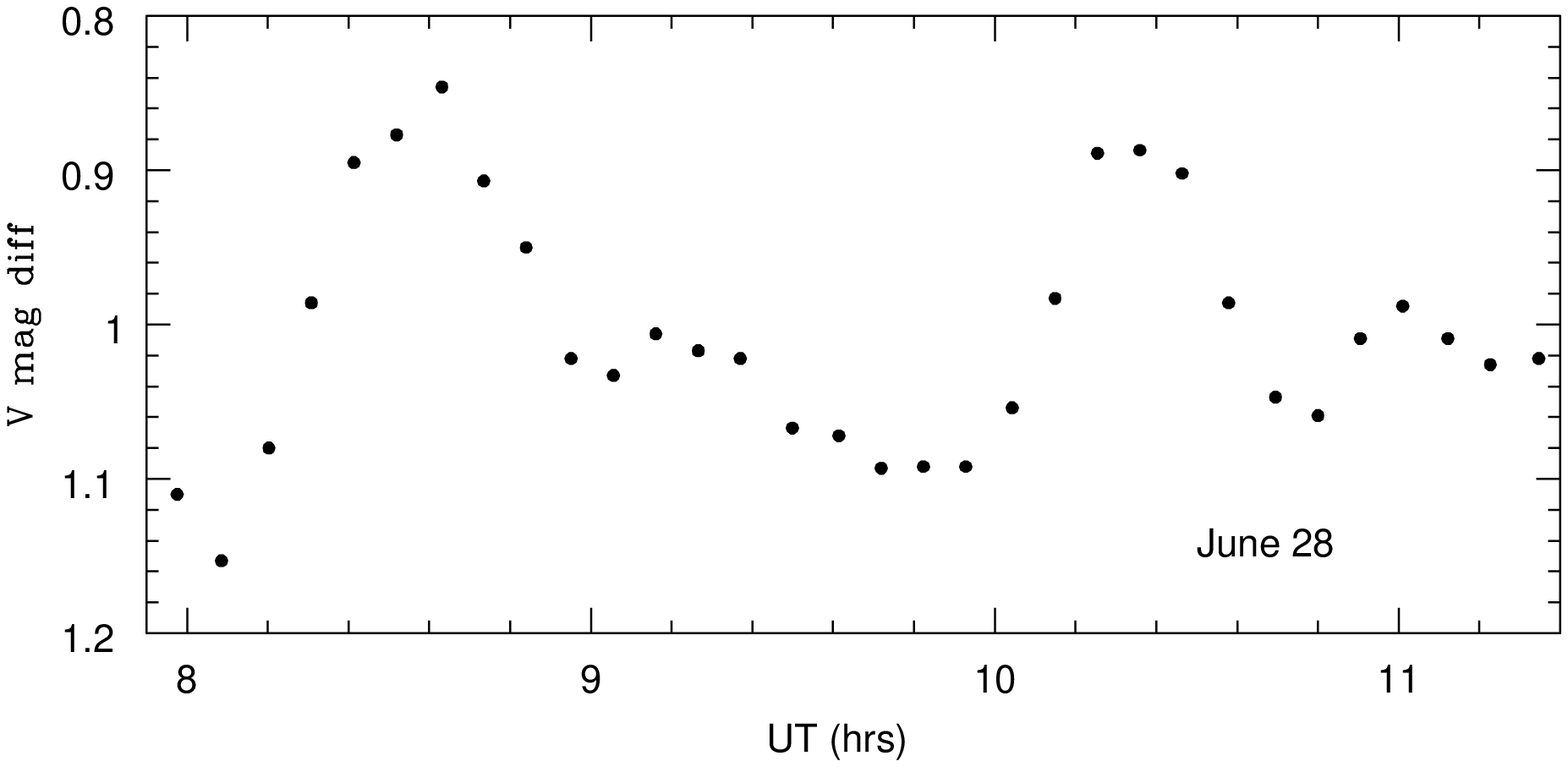}
\caption{KPNO light curve of the dwarf nova CSS/SDSS2158+09 obtained 2011 June 28, showing a possible orbital period near 104 min with hints of eclipses.}
\end{figure}

\clearpage
\begin{figure}
\figurenum {12}
\plotone{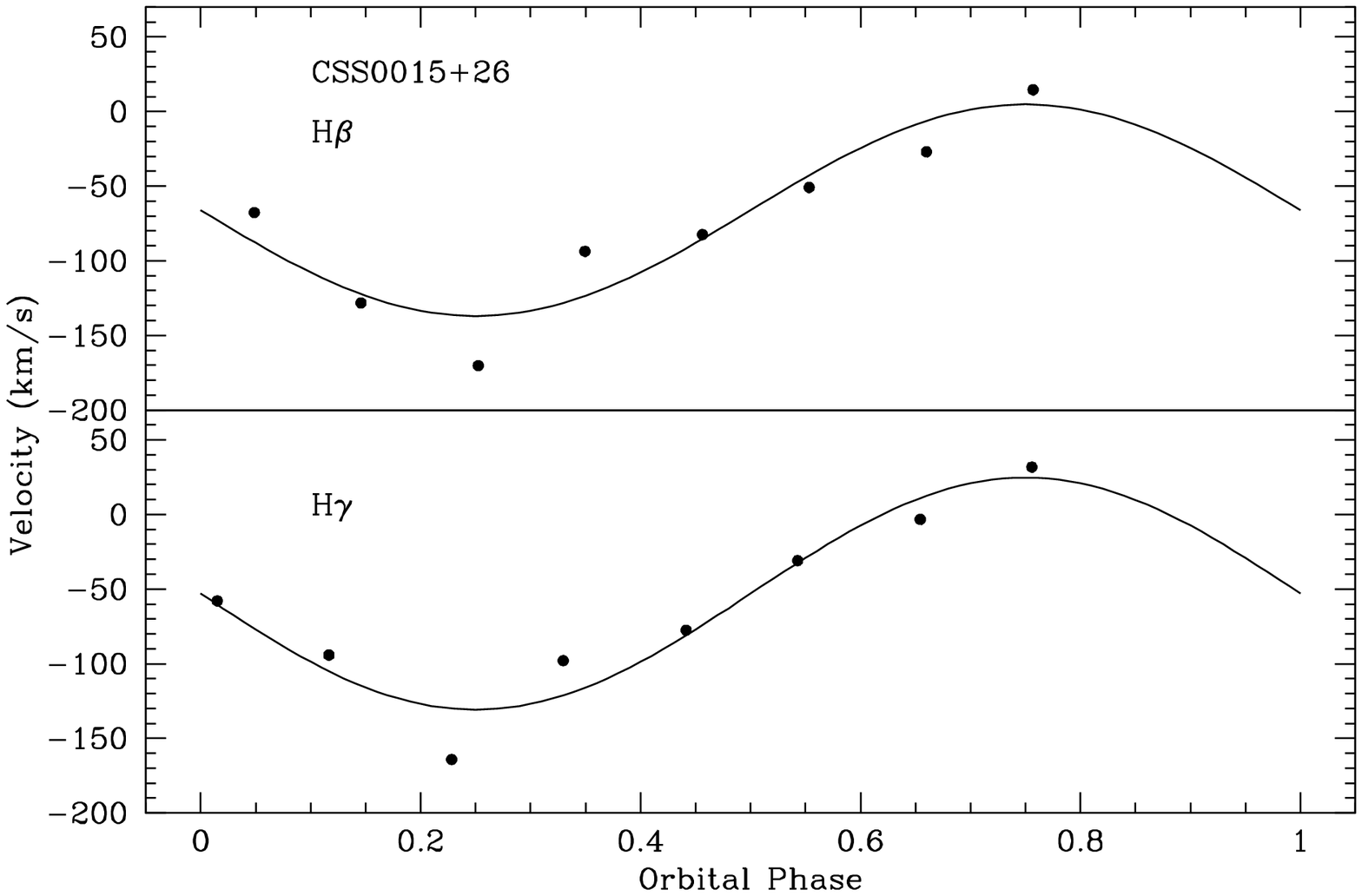}
\caption{Velocity curves of H$\beta$ and H$\gamma$ obtained for the dwarf nova 
CSS0015+26 on 
2011 August 26. H$\beta$ is phased with a period of 103 min and H$\gamma$ with
a period of 99 min. Best fit sinusoids are shown with parameters listed in
Table 3.}
\end{figure}

\clearpage
\begin{figure}
\figurenum {13}
\plotone{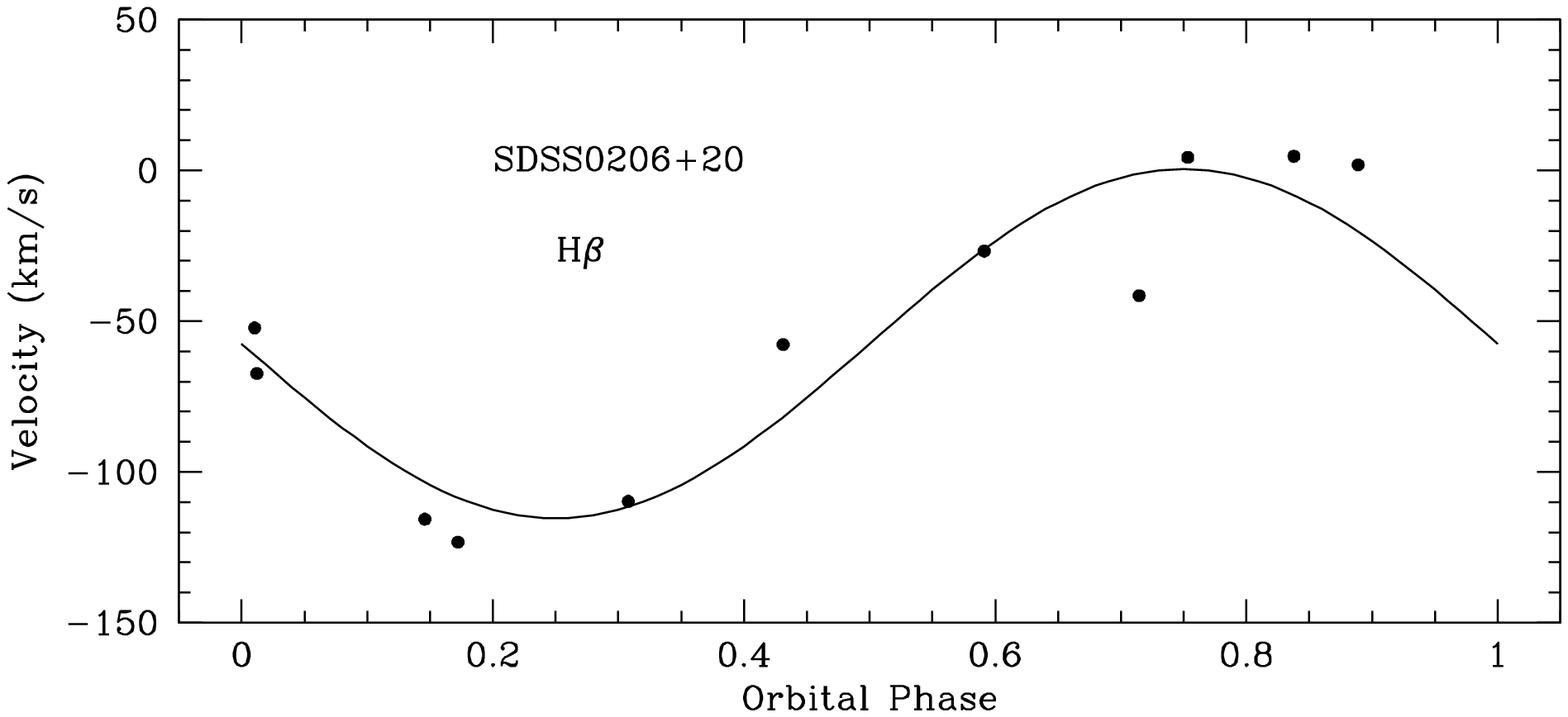}
\caption{H$\beta$ velocity curve of the dwarf nova CSS/SDSS0206+20 obtained on 
2013 September 6, phased with a period of 81 min and showing the best fit
sine curve with parameters from Table 3.}
\end{figure}

\clearpage
\begin{figure}
\figurenum {14}
\plotone{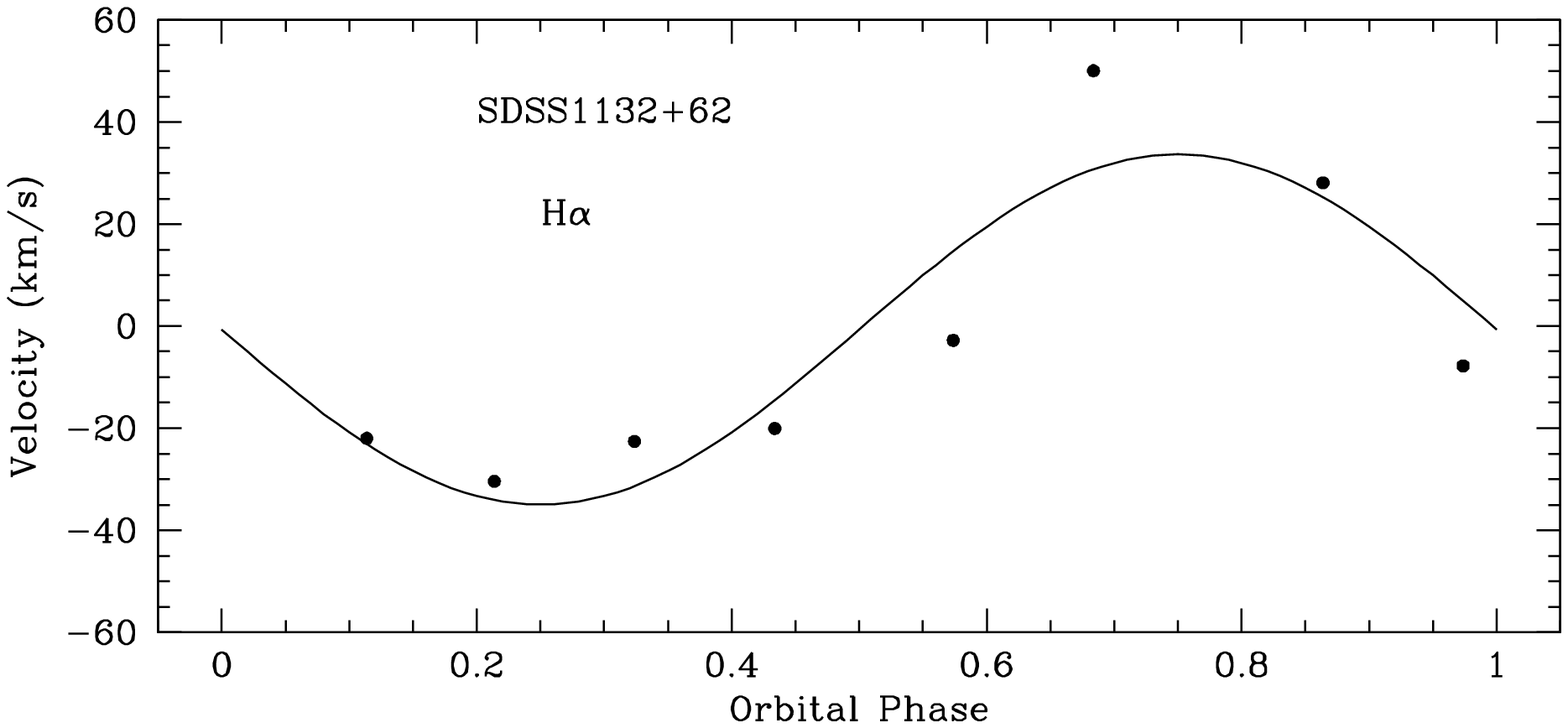}
\caption{H$\alpha$ velocity curve of the dwarf nova SDSS1132+62 obtained on 
2013 February 5, phased with a period of 100 min and the best fit sine curve
with parameters from Table 3.}
\end{figure}
 

\begin{thebibliography}{}

\bibitem[Breedt et al. (2014)]{B14} Breedt, E. et al. 2014, \mnras, in press (B14)

\bibitem[Bruch \& Schimpke (1992)]{Br92} Bruch, A. \& Schimpke, T. 1992, \aaps, 93, 419

\bibitem[Coppejans et al. (2014)]{C14} Coppejans, D. L. et al. 2014, \mnras, 437, 510

\bibitem[Denisenko (2013a)]{D13a} Denisenko, D. V. 2013a, vsnet-alert 16406

\bibitem[Deniskenko (2013b)]{D13b} Denisenko, D. V. 2013b, vsnet-alert 16457

\bibitem[Drake et al. (2009a)]{Dr9a} Drake, A. J. et al. 2009a, \apj, 696, 870

\bibitem[Drake et al. (2009b)]{Dr9b} Drake, A. J. et al. 2009b, ATEL, 2210

\bibitem[G\"ansicke et al. (2009)]{G09} G\"ansicke, B. T. et al. 2009, \mnras, 397, 2170

\bibitem[Green et al. (1982)]{Gr82} Green, R. F., Ferguson, D. H., Liebert, J. \& Schmidt, M. 1982, \pasp, 94, 560 

\bibitem[Hagen et al. (1995)]{H95} Hagen, H. J., Groote, D., Engels, D. \& Reimers, D. 1995, \aaps, 111, 195

\bibitem[Kato (2013a)]{K13a} Kato, T. 2013a, vsnet-alert 15803

\bibitem[Kato (2013b)]{K13b} Kato, T. 2013b, vsnet-alert 16302

\bibitem[Liebert et al. (1978)]{L78} Liebert, J., Stockman, H. S., Angel, J.R. P., et al. 1978 \apj, 225, 201

\bibitem[Lipunov et al. (2010)]{Li10} Lipunov, V. M. et al. 2010, AdAst, 349171 

\bibitem[Maehara (2011)]{M11} Maehara, H. 2011, vsnet-alert 13412

\bibitem[Miller (1971)]{Mi71} Miller, W. J. 1971, Spec. Vat. Ric. Astron. 8, 167

\bibitem[Oshima (2011a)]{O11a} Oshima, T. 2011a, vsnet-alert 13289

\bibitem[Oshima (2011b)]{O11b} Oshima, T. 2011b, vsnet-alert 13357

\bibitem[Oshima (2013)]{O13} Ohshima, T. 2013, vsnet-alert 16294

\bibitem[Pojmanski (1997)]{P97} Pojmanski, G. 1997, AcA, 47, 467

\bibitem[Stanek (2013)]{S13} Stanek, K. 2013, vsnet-alert 16272

\bibitem[Szkody et al. (2004)]{Sz04} Szkody, P. et al. 2004, \aj, 128, 1882

\bibitem[Szkody et al. (2006)]{Sz06} Szkody, P. et al. 2006, \aj, 131, 973

\bibitem[Szkody et al. (2009)]{Sz09} Szkody, P. et al. 2009, \aj, 137, 4011

\bibitem[Szkody et al. (2011)]{Sz11} Szkody, P. et al. 2011, \aj, 142, 181

\bibitem[Thorne et al. (2010)]{T10} Thorne, K., Garnavich, P. \& Mohrig, K. 2010, IBVS 5923

\bibitem[Thorstensen \& Skinner (2012)]{TS12} Thorstensen, J. R. \& Skinner, J. N. 2012, \aj, 144, 81 (TS)

\bibitem[Warner(1995)]{W95}
Warner,~B., 1995, Cataclysmic Variable Stars [Cambridge: University Press]

\bibitem[Warner \& Nather (1972)]{WN72} Warner, B. \& Nather, R. E. 1972, \mnras, 156, 305

\bibitem[Wheatley \& Ramsay (1998)]{WR98} Wheatley, P. J. \& Ramsay, G. 1998, ASPCS, 137, 446

\bibitem[Wickramasinghe \& Ferrario (2000)]{WF00} Wickramasinghe, D.T. \& Ferrario, L. 2000, \pasp, 112, 873

\bibitem[Woudt et al. (2012)]{W12} Woudt, P. A., Warner, B., de Bude, D., et al. 2012, \mnras, 421, 2414

\bibitem[York et al. (2000)]{Y00} York, D. G., Adelman, J., Anderson, J. E., Jr., et al. 2000, \aj, 120, 1579


\end{thebibliography}
\end{document}